\documentclass[pdflatex,sn-mathphys-num]{sn-jnl}

\usepackage{graphicx}%
\usepackage{multirow}%
\usepackage{amsmath,amssymb,amsfonts}%
\usepackage{amsthm}%
\usepackage{mathrsfs}%
\usepackage[title]{appendix}%
\usepackage{xcolor}%
\usepackage{textcomp}%
\usepackage{manyfoot}%
\usepackage{booktabs}%
\usepackage{algorithm}%
\usepackage{algorithmicx}%
\usepackage{algpseudocode}%
\usepackage{listings}%
\usepackage{multicol} 
\usepackage{subcaption}
\usepackage{hyperref}
\usepackage{pifont} 
\usepackage{array}
\usepackage[labelfont=bf]{caption}  

\theoremstyle{thmstyleone}%
\theoremstyle{thmstyletwo}%

\theoremstyle{thmstylethree}%

\begin{document}

\title[Article Title]{Assess Space-Based Solar Power in European-Scale Power System Decarbonization}


\author[1,2]{\fnm{Xinyang} \sur{Che}}

\author[3]{\fnm{Lijun} \sur{Liu}}

\author*[1]{\fnm{Wei} \sur{He}}\email{wei.4.he@kcl.ac.uk}

\affil*[1]{\orgdiv{Department of Engineering}, \orgname{King's College London}, \orgaddress{ \city{London}, \country{UK}}}

\affil[2]{\orgdiv{School of Physics}, \orgname{Xi'an Jiaotong University}, \orgaddress{\city{Xi'an}, \country{China}}}

\affil[3]{\orgdiv{School of Energy and Power Engineering}, \orgname{Xi'an Jiaotong University}, \orgaddress{\city{Xi'an}, \country{China}}}


\abstract{

Meeting net-zero targets remains formidable as terrestrial renewables grapple with intermittency and regional variability. Here, we integrate space-based solar power—a potential near-constant, orbital solar technology—into a high-resolution, Europe-wide capacity-expansion and dispatch model to quantify its contribution under net-zero constraints. We examine two advanced space solar designs: (1) a near-baseload, low technology readiness level concept (heliostat-based representative design one) and (2) a partially intermittent, higher technology readiness level concept (planar-based representative design two), both drawing on NASA’s 2050 cost and performance projections. Our results show that heliostat-based design can reduce total system costs by 7–15\%, displace up to 80\% of intermittent wind and solar, and cut battery usage by over 70\%, if it meets its forecast cost reductions—though long-duration storage (e.g., hydrogen) remains essential for seasonal balancing. By contrast, planar-based design is economically unattractive at its projected 2050 costs. Through extensive sensitivity analyses, we identify cost thresholds at which space-based solar power shifts from cost-prohibitive to complementary and ultimately to a dominant baseload technology. Specifically, heliostat-based design becomes complementary at roughly 14× and dominant at 9× the 2050 solar PV capital cost, benefiting from its continuous power generation. Meanwhile, planar-based design must achieve even lower cost levels (9× to be complementary and 6× to dominate) and would rely on short-duration storage to mitigate its partial intermittency. These findings provide quantified techno-economic benchmarks and reveal alternative net-zero pathways, offering critical guidance for policymakers and industry stakeholders seeking large-scale, centrally coordinated renewable solutions with non- or low-intermittency.


}




\maketitle

\section{Background} 



By November 2023, nearly 145 countries, accounting for about 90\% of global greenhouse gas emissions, have pledged to reach net-zero targets \cite{netzero_countries}. Meeting these commitments requires shifting from fossil fuels to low-carbon resources such as wind and solar power. Yet, the intermittency and weather-dependence of these terrestrial renewables complicate reliable supply, and the cost-effectiveness of large-scale, long-duration storage remains uncertain, challenging deep decarbonization \cite{schmidt2019projecting, he2021technologies}.

In Europe, decarbonization efforts take place within a complex, highly interconnected grid spanning diverse regional resources, demand profiles, and policies. Seasonal imbalances (e.g., higher winter demand met significantly by natural gas) can spur price volatility, emissions, and energy security risks \cite{mivsik2022eu}. Whole-system optimization models indicate that achieving high renewable penetration is both technologically and economically feasible \cite{victoria2020early}, but such solutions involve intricate coordination of generation, storage, and extensive transmission upgrades across national borders. As policymakers strive for robust low-carbon pathways, identifying complementary options that mitigate intermittency without overwhelming grid infrastructure remains a pressing challenge.

Space-based solar power (SBSP) could offer a centralized, weather-independent energy resource to help address these challenges \cite{glaser1974feasibility}. By operating above the atmosphere and outside the day-night cycle, SBSP promises continuous gigawatt-scale power. Recent technological milestones suggest it may evolve from a niche concept to a techncially viable solution by the 2030s: multi-junction and lightweight PV cells achieving near 47\% efficiency \cite{geisz2020six, lang2020proton}, modular in-orbit assembly \cite{spiderfab}, and successful wireless power demonstrations \cite{ayling2024wireless, pelham2024lyceanem} have all reached mid-range Technology Readiness Levels (TRL 4--5). Meanwhile, reusable launch vehicles have significantly reduced launch costs, with some estimates projecting a Levelized Cost of Energy (LCOE) for SBSP of \$30--80\,MWh$^{-1}$ by 2050 \cite{jones2018recent, osoro2021techno, rodgers2024space, mizrahi2024space}. In parallel, advances in integrated uncertain optimal design for truss configurations \cite{yang2025integrated}, interval Riccati equation-based dynamic analysis \cite{YANG2025118742}, convex set reliability-based optimal attitude control \cite{YANG2025115769}, and multi-objective optimization for robust attitude determination \cite{YANG20243273} have further strengthened the technical foundations of SBSP. Although considerable uncertainties remain—from in-orbit manufacturing to policy frameworks—major space agencies (NASA, ESA, JAXA) are actively shaping regulatory pathways, motivating the need to understand SBSP’s potential contribution to net-zero goals \cite{rodgers2024space, ESA_sbsp, JAXA}.

From the implementation perspective, Europe’s longstanding tradition of multinational cooperation—including cross-border electricity exchange and satellite ventures under ESA—could be leveraged to develop and operate a centralized SBSP infrastructure. As a continent-scale solution to provide stable, baseload-scale renewable supply, SBSP would reduce the continent’s reliance on gas-fired power, thereby lowering emissions and enhancing energy security.

Despite these developments and opportunities, most SBSP research focuses on technical feasibility rather than on integration within a future energy mix dominated by terrestrial renewables. This leaves a critical gap: determining when and where SBSP might be cost-effective or synergistic with storage and expanded transmission. Given long lead times for infrastructure planning, these questions merit immediate study to guide R\&D priorities, market evolution, and policy design.

To address this gap, our study pursues three key objectives:
\begin{itemize}
    \item {Integrate SBSP into a continental-scale model}: We embed SBSP into a European-scale capacity-expansion and dispatch framework, focusing on meeting electricity demand as defined by ENTSO-E’s 2050 native demand \cite{TYNDP2024}, and evaluate its interactions with terrestrial renewables, storage, and transmission infrastructure under net-zero constraints.
    \item {Assess grid balancing benefits}: We quantify how continuous SBSP can reduce storage requirements and transmission investments by offsetting the temporal and spatial variability of terrestrial wind and solar.
    \item {Identify techno-economic conditions}: We determine cost thresholds at which SBSP becomes competitive, providing tangible guidance for research, policy, and investment decisions.
\end{itemize}

We employ scenario- and sensitivity-based methods to capture uncertainties in SBSP costs, efficiencies, and operational factors, ensuring robust, policy-relevant insights. Rather than advancing SBSP technology in isolation, this analysis clarifies how SBSP might complement or compete with existing decarbonization pathways. By identifying system-level impacts and techno-economic trade-offs, we offer actionable information for policymakers, industry leaders, and researchers seeking to shape Europe’s evolving clean energy landscape.

\section{SBSP Concept and Generation Modeling}

\subsection{SBSP Concept}

\begin{figure}[htbp]
    \centering
    \begin{subfigure}[b]{0.9\textwidth}
        \includegraphics[width=\textwidth]{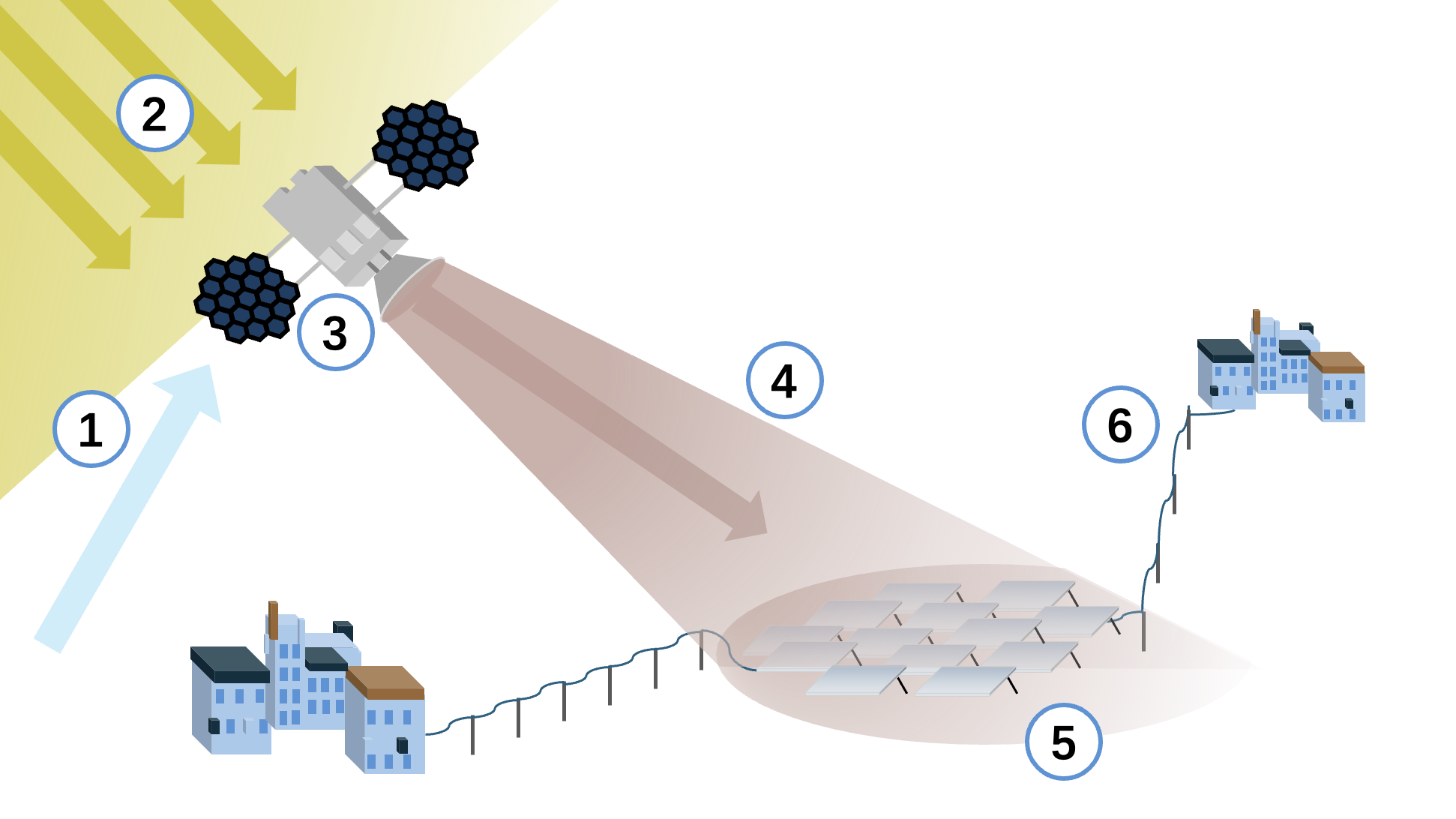}
        \caption{}
        \label{fig:sbsp_a}
    \end{subfigure}
    

    \begin{subfigure}[b]{0.9\textwidth}
        \includegraphics[width=\textwidth]{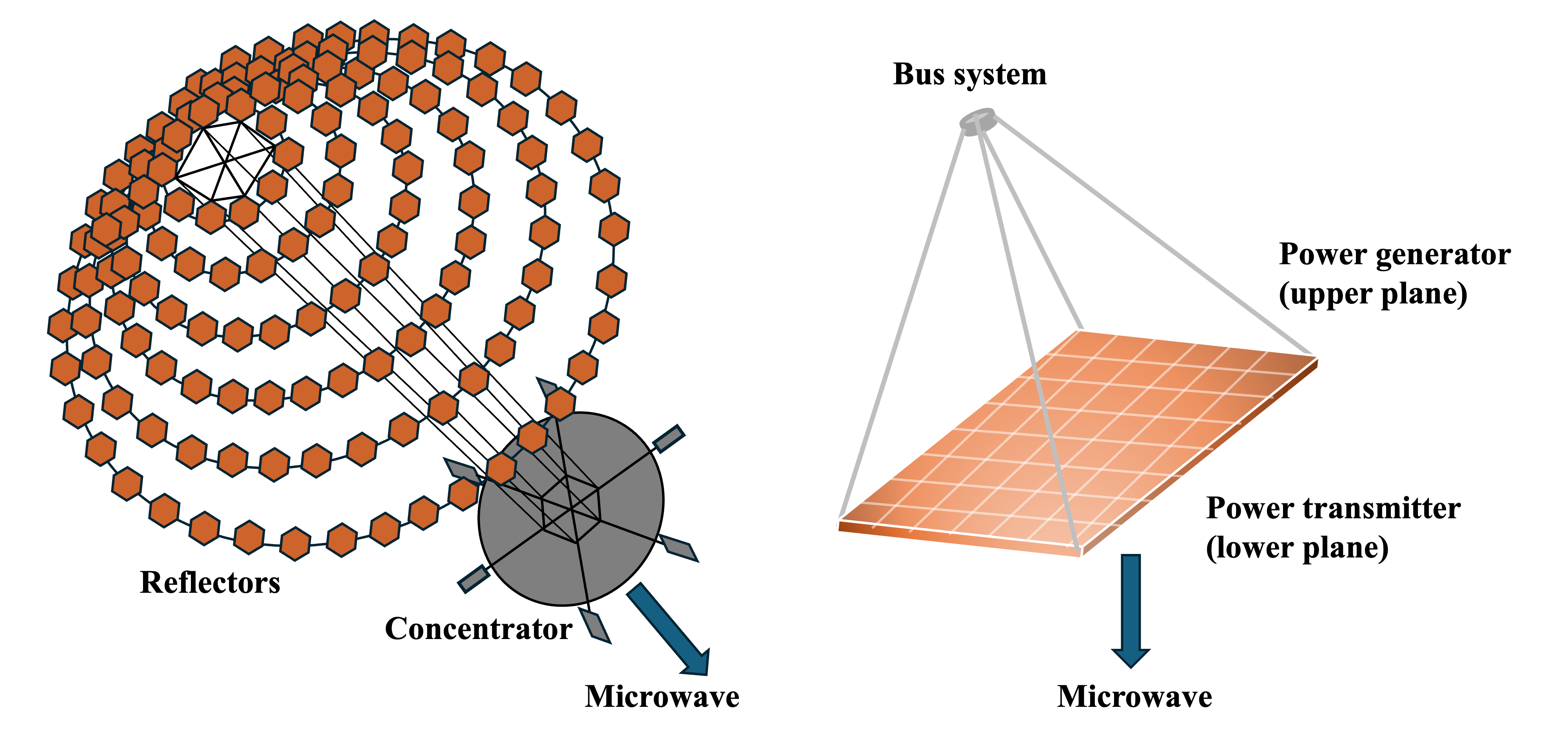}
        \caption{}
        \label{fig:sbsp_b}
    \end{subfigure}
    
    \caption{\textbf{Overview of SBSP Operational Process and System Architectures.} 
    \textbf{(a)} Stepwise operational process of a SBSP system, including: (1) launch and installation in space, (2) solar energy collection, (3) conversion to electricity and then to microwave, (4) transmission to Earth, (5) reception and reconversion on the ground, and (6) grid delivery. 
    \textbf{(b)} Main structures of two representative SBSP system designs. The left panel shows the Innovative Heliostat Swarm concept, which is broadly derived from the Alpha Mark III architecture \cite{ref13}. This design employs reflectors and a central concentrator to continuously focus sunlight throughout the day, using independently operating hexagonal modules arranged in a beehive-like configuration. The right panel depicts the mature Planar Array system with a sandwich architecture: solar collection on one side and microwave transmission on the other. Identical power modules convert solar energy to microwave power and are wirelessly controlled by a central bus, while Earth-facing antennas maintain orientation via gravity gradient forces in GEO.}
    \label{fig:sbsp_process_designs}
\end{figure}

Figure~\ref{fig:sbsp_a} illustrates the main stages of an SBSP system, from satellite deployment in orbit to grid integration on Earth. Most designs target geostationary Earth orbit (GEO), where continuous solar exposure enables stable, high-capacity power generation with minimal eclipse times \cite{ref1}. After launch, robotic systems assemble modular components, such as lightweight photovoltaic (PV) arrays or concentrating mirrors, for solar collection \cite{ref2, ref3}. The harvested electricity is then converted into microwave or laser beams for transmission. Microwave frequencies (1--10\,GHz), especially 2.45\,GHz or 5.8\,GHz \cite{ref4}, are commonly selected to balance transmission efficiency, atmospheric attenuation, and safety constraints \cite{ref5, ref6}.

On the ground, rectifying antennas (rectennas) spanning several square kilometers capture the transmitted energy and convert it to direct current (DC), which is then inverted to alternating current (AC) and fed into the existing grid infrastructure. Lightweight rectenna designs allow partial land co-use, and multiple stations can share power from a single satellite \cite{ref1}. A centralized control center manages satellite pointing, beam transmission, and diagnostics, ensuring real-time coordination and leveraging GEO’s continuous solar availability to offer a high-capacity, dispatchable renewable energy source.

Although originally conceived by Peter Glaser in 1968 \cite{ref7}, SBSP remained impractical in early evaluations due to prohibitively high costs for launch, in-orbit assembly, and maintenance \cite{ref8, ref9}. Over the past three decades, however, significant advances in solar cell efficiency, wireless power transmission, and reusable launch vehicles have reignited interest in SBSP. A variety of novel designs have since emerged, including 
the Large Array of Solar Panels \cite{ref7}, Sun Tower Design \cite{ref9}, Modular Symmetrical Concentrator \cite{Modular}, Solar Power Satellite via Arbitrarily Large Phased Array \cite{Arbitrarily}, Space-Based Reflectors \cite{reflectors}, Inflatable Thin-Film Structures \cite{thin-film}, Hypermodular SBSP Design \cite{Hypermodular}, and Tethered Satellite System \cite{Tethered}. Each concept seeks to optimize power density, structural mass, and reliability under the demanding conditions of orbit (Details in the Supplementary Information Section 4).

Within this landscape, NASA recently proposed two representative architectures that stand out for their integrative approach, incorporating multiple previously tested or actively developing subsystems into cohesive, up-to-date SBSP designs \cite{ref11}:

\begin{itemize}
    \item \textbf{Representative Design One (RD1), the Innovative Heliostat Swarm:} 
    Based on the SPS-ALPHA Mark~III concept \cite{ref10, ref13, ref14}, this lower-TRL design uses mirror-like reflectors (heliostats) to direct sunlight to a central concentrator, enabling nearly 99.7\% annual power availability \cite{ref10}. Key innovations include retro-directive RF transmission arrays, high-efficiency PV cells, lightweight modular structures, and autonomous in-orbit deployment. Figure~\ref{fig:sbsp_b} shows how reflectors continuously adjust orientation, maximizing solar capture throughout the orbital cycle.
    
    \item \textbf{Representative Design Two (RD2), the Mature Planar Array:} 
    Adapted from JAXA’s Tethered Solar Power Satellite \cite{ref3} and Caltech’s SSPP design \cite{ref12}, planar design (RD2) includes planar panels whose lower surface faces Earth under gravity-gradient forces. While solar incidence on the upper and lower surfaces varies with orbital geometry, the system achieves roughly 60\% annual power availability \cite{ref3}. Figure~\ref{fig:sbsp_b} illustrates its higher TRL, given demonstrated hardware and well-documented performance characteristics.

\end{itemize}

These two concepts differ in TRL and technical approach yet both incorporate high-efficiency solar capture, wireless power transmission, and modular architectures designed to lower launch mass and total costs. By examining heliostat design (RD1, lower TRL) and planar design (RD2, higher TRL) side-by-side, we provide a balanced assessment of SBSP performance, feasibility, and cost. This approach captures essential trade-offs in orbital assembly complexity, antenna design, and seasonal power availability, ensuring our analysis thoroughly reflects the technical and economic considerations that will guide future SBSP deployment.

\subsection{SBSP Modeling}
To assess the system-level impacts of SBSP, we modeled the generation profiles and lifetime costs of NASA’s two representative designs, heliostat design (RD1) and planar design (RD2), for both current (2020) and future (2050) scenarios. Drawing on published technical data \cite{ref11}, we derived hourly generation outputs by accounting for orbital geometry, eclipse periods, solar incidence angles, and design-specific operational constraints (e.g., RD1’s heliostat-based approach vs.\ RD2’s planar configuration). We assume the geostationary orbit altitude (\(\sim 35{,}786\text{ km}\)) is negligible relative to the Earth–Sun distance (\(\sim 1.49\times10^8\text{ km}\)) \cite{gurnett2009search,ref11}. Complete derivations and assumptions, including the inverse-square law for irradiance and eclipse scheduling (detailed in the Supplementary Information Section~1).

\begin{figure}[htbp]
    \centering
    \begin{subfigure}[t]{0.7\textwidth}
        \centering
        \raisebox{0.5\height}{\textbf{(a)}}\hspace{-0.5cm}
        \includegraphics[width=\textwidth]{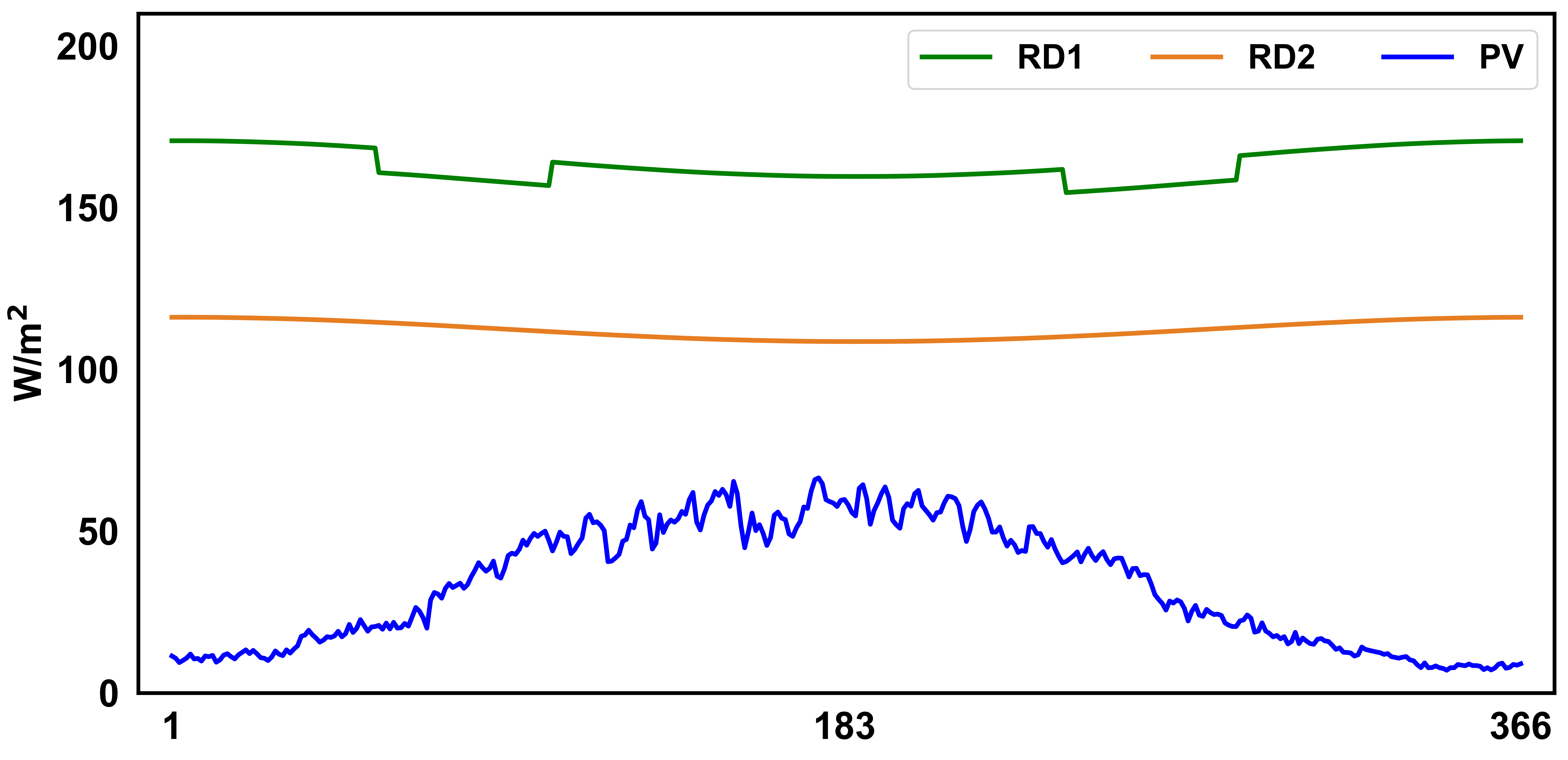}
        \label{fig:comparison_pv_day}
    \end{subfigure}

    \vspace{0.5cm}

    \begin{subfigure}[t]{0.9\textwidth}
        \centering
        \raisebox{0.5\height}{\textbf{(b)}}\hspace{-0.5cm}
        \includegraphics[width=\textwidth]{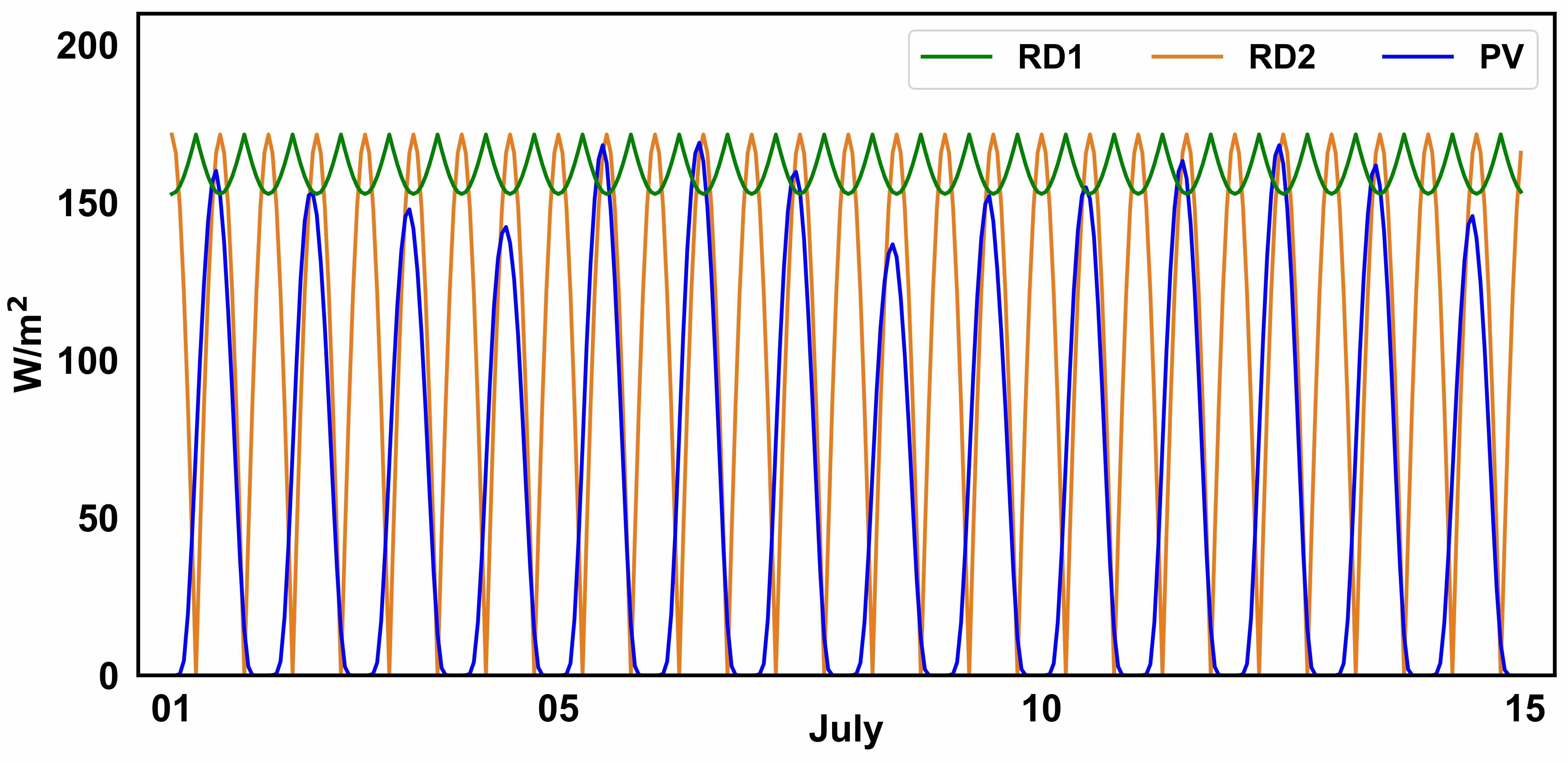}
        \label{fig:comparison_pv_2}
    \end{subfigure}

    \vspace{0.5cm} 

    \begin{subfigure}[t]{0.9\textwidth}
        \centering
        \raisebox{0.5\height}{\textbf{(c)}}\hspace{-0.5cm}
        \includegraphics[width=\textwidth]{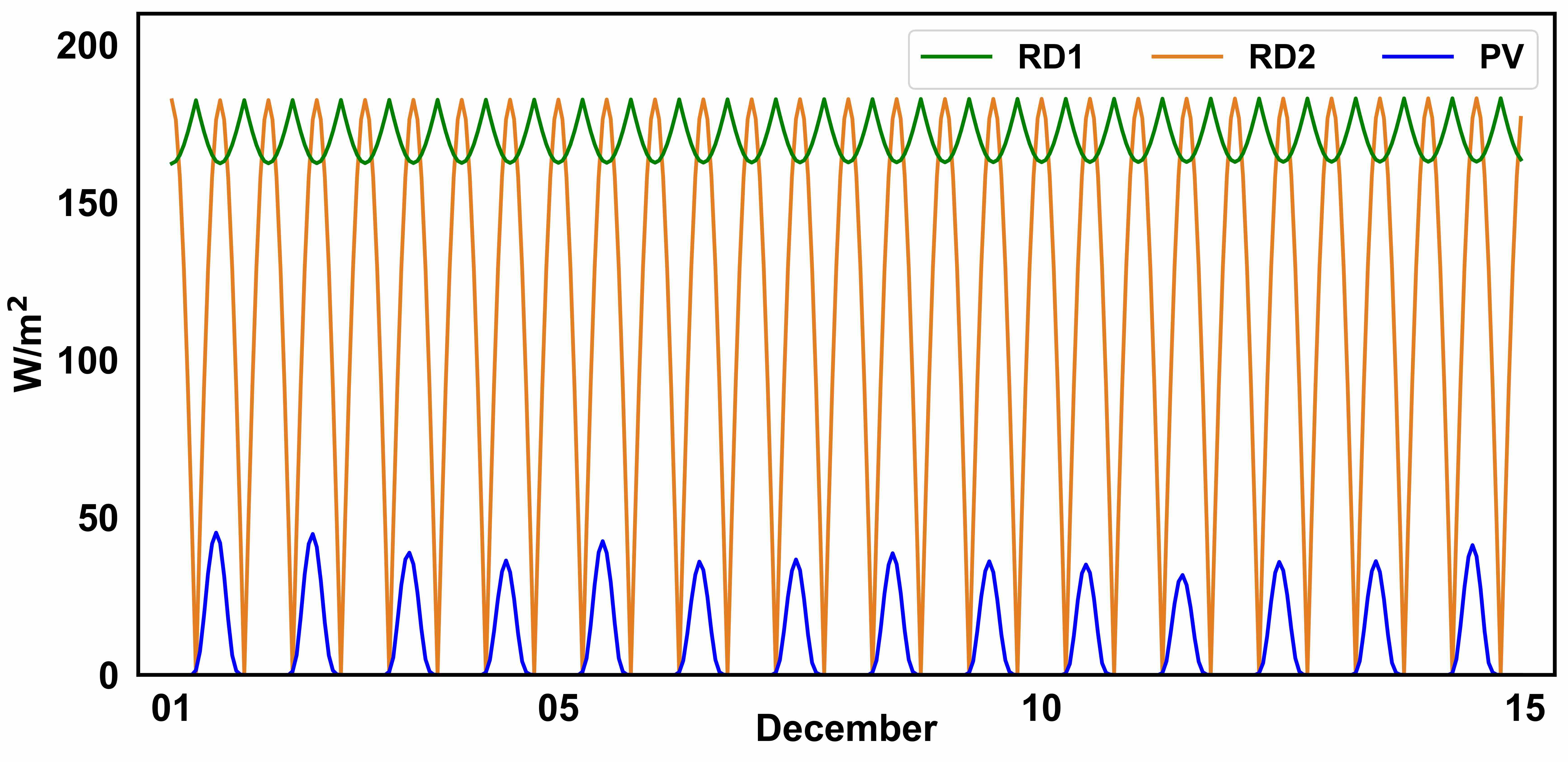} 
        \label{fig:comparison_pv_1}
    \end{subfigure}

    \caption{\textbf{Comparison of the variations in power output per square meter of heliostat design (RD1), planar design (RD2), and terrestrial solar panels in 2020.} (a) To clarify the power generation curves of heliostat design, planar design, and terrestrial solar, the hourly data were averaged to obtain daily mean values, enabling the observation of overall trends throughout 2020. (b) Hourly data from July 1 to July 15 represent the summer period. (c) Hourly data from December 1 to December 15 represent the winter period.}
    \label{fig:comparison_pv}
\end{figure}

\vspace{0.5em}
\noindent\textbf{Generation Profiles.}
Both heliostat design (RD1) and planar design (RD2) capture solar energy in GEO nearly year-round, but differ in achievable annual availability: heliostat design uses heliostats to concentrate sunlight, enabling up to 99.7\% power availability, while planar design’s planar panels achieve approximately 60\% due to geometric constraints \cite{ref10,ref3}. Compared to terrestrial solar panels whose power availability is usually 15-30\% \cite{Lazard2023LCOE}, both heliostat design and planar design show improved power availability. We simulate and compare the hour-by-hour power output per square meter of heliostat design, planar design, and terrestrial solar panels, based on representative orbital parameters, solar irradiation, panel characteristics, and efficiencies (SI, Section 1). Using 2020 as a case study, the results shown in Figure~\ref{fig:comparison_pv} demonstrate the potential advantages of SBSP power generation. The power availability of heliostat design and planar design exceeds that of terrestrial solar by 70\% and 42\%, respectively. Furthermore, their power output is more continuous, significantly improving efficiency, especially for heliostat design. During the continuous phase, the power output per square meter of heliostat design and planar design is higher at different times compared to terrestrial solar. Over the course of 2020, the power output per square meter of heliostat design and planar design panels is 384\% and 233\% higher than that of terrestrial solar, respectively. Terrestrial solar has higher output in summer than in winter, while SBSP exhibits the opposite trend, providing complementary power generation in the future. Additionally, the power output of SBSP is significantly less affected by seasonal changes compared to terrestrial solar, indicating its potential for more stable power generation with smaller fluctuations.

\vspace{0.5em}
\noindent\textbf{Cost Analysis.}
We adopt NASA’s five-phase concept of operations (ConOps) framework (Develop, Assemble, Operate, Maintain, and Dispose) to estimate capital expenditures (CapEx) and fixed operation and maintenance costs (FOM) \cite{ref11}. The \emph{Develop} and \emph{Assemble} phases, covering technology R\&D, hardware manufacturing, launch, and in-orbit assembly, are aggregated into CapEx, while \emph{Operate}, \emph{Maintain}, and \emph{Dispose} phases contribute to FOM. Key cost drivers include reusable launch vehicles, modular assembly robotics, and lifetime maintenance. Variable Operations and Maintenance (VOM) costs typically apply to electricity fuel sources. However, since SBSP does not rely on fuel as a direct input for electricity production, this cost category, along with Fuel Costs, is excluded for all renewable electricity production technologies \cite{ref11}. Under baseline assumptions, we find that near-term (2020) SBSP costs are high, but can drop by 2050 with NASA's projected efficiencies, launch cost reductions, and technological improvements (SI, Section~2). Building on this foundation, we conducted multiple variable sensitivity analyses, incorporating predicted data for various indicators based on NASA’s projections for 2050 \cite{ref11}. The estimated costs of heliostat design and planar design in 2020 (the baseline case) and 2050 are shown in Table \ref{tab:costs}. CapEx refers to the capital expenditures required to acquire, upgrade, or maintain long-term assets, such as equipment, while FOM represents the annual costs associated with the fixed operation and maintenance of these assets. VOM, on the other hand, refers to the variable operation and maintenance costs that change with energy production or system usage \cite{nrel2022definitions}.

To capture total system costs comprehensively, we represent each technology in the capacity-expansion model with two cost components: {annualized capital costs} and {marginal costs}. We convert the {CapEx} (investment cost) values for SBSP into {annualized capital costs} (including fixed O\&M) using the annuitization approaches and exchange rates in \cite{horsch2018pypsa}, as summarized in Table~\ref{tab:costs}. Marginal costs cover variable O\&M (VOM), plus any additional expenses (e.g., fuel) that arise when output increases by one unit. By incorporating the uncertainties in orbital efficiency, beam transmission losses, and cost trajectories, we evaluate roles of SBSP for net zero under different future scenarios. Further details on modeling equations, sensitivity parameters, and data sources are included in the SI. In the subsequent results of this paper, all references to capital costs and marginal costs correspond to the data used in the model. 

\begin{table}[h!]
\centering
\renewcommand{\arraystretch}{1.5} 
\setlength{\tabcolsep}{10pt} 
\caption{Key Parameters \cite{ref11} and Costs in models (PyPSA-Eur) for SBSP under baseline and 2050 scenarios.}
\begin{tabular}{p{2cm}cccc} 
\toprule
\textbf{SBSP} & \textbf{Lifetime (Years)} & \textbf{Output (MW)} & \textbf{Capacity Factor (\%)} & \textbf{Discount Rate (\%)} \\
\midrule
Heliostat Baseline & 30 & 2,028.79 & 99.7 & 3 \\
Planar Baseline & 30 & 2,021.95 & 60 & 3 \\
Heliostat 2050 & 30 & 2,028.79 & 99.7 & 3 \\
Planar 2050 & 30 & 2,021.95 & 60 & 3 \\
\bottomrule
\end{tabular}

\vspace{0.2cm} 

\begin{tabular}{cccc} 
\toprule
\textbf{CapEx (EUR/kW)} & \textbf{FOM (EUR/kW-year)} & \textbf{Capital cost (EUR/kW$\cdot$year)} & \textbf{Marginal cost (EUR/kW)} \\
\midrule
41,071.35 & 2,806.24 & 4,901.65 & 0 \\
62,335.91 & 4,506.15 & 7,687.48 & 0 \\
2,915.31 & 119.15 & 267.87 & 0 \\
3,701.30 & 207.71 & 396.59 & 0 \\
\bottomrule
\end{tabular}
\label{tab:costs}
\end{table}

\section{Results}

\subsection*{Interaction with terrestrial renewables in 2050 power systems}

We individually integrate heliostat design (RD1) and planar design (RD2) into the PyPSA-Eur model for minimum-cost optimization under both current (2020) and future (2050) scenarios (see “Methods”). The results for the current scenario are provided in SI Section 5, as neither heliostat design nor planar design is selected in the optimized system. This outcome highlights the economic infeasibility of SBSP under present-day conditions due to its high capital cost. We evaluate the value and impacts of SBSP across cost, temporal, and spatial dimensions under three main 2050 scenarios. These scenarios—termed low, medium, and high SBSP market opportunity—differ in the capital costs of available generation resources, which include solar photovoltaics (PV), onshore and offshore wind, reservoir hydro, biomass, run-of-river hydro, and nuclear. The capital costs of these resources, listed in Table~\ref{tab:capital_cost}, are lowest in the low-opportunity scenario and highest in the high-opportunity scenario.


\begin{table}[ht]
\centering
\caption{\textbf{Capital costs in EUR/kW$\cdot$year in 2050, for the three market opportunity scenario classes}}
\begin{tabular}{lccc}
\toprule
\textbf{} & \textbf{Low} & \textbf{Medium} & \textbf{High} \\
\midrule
Solar & 27.68 & 32.29 & 41.74 \\
Onshore wind & 72.36 & 103.75 & 121.46 \\
Offshore wind & 197.30 & 211.70 & 222.12 \\
Reservoir & 216.67 & 216.67 & 216.67 \\
Biomass & 226.31 & 229.89 & 233.40 \\
Run-of-river & 318.33 & 319.88 & 321.43 \\
Nuclear & 443.76 & 443.76 & 443.76 \\
\bottomrule
\end{tabular}
\label{tab:capital_cost}
\end{table}

\begin{figure}[htbp]
    \centering
    \includegraphics[width=1\textwidth]{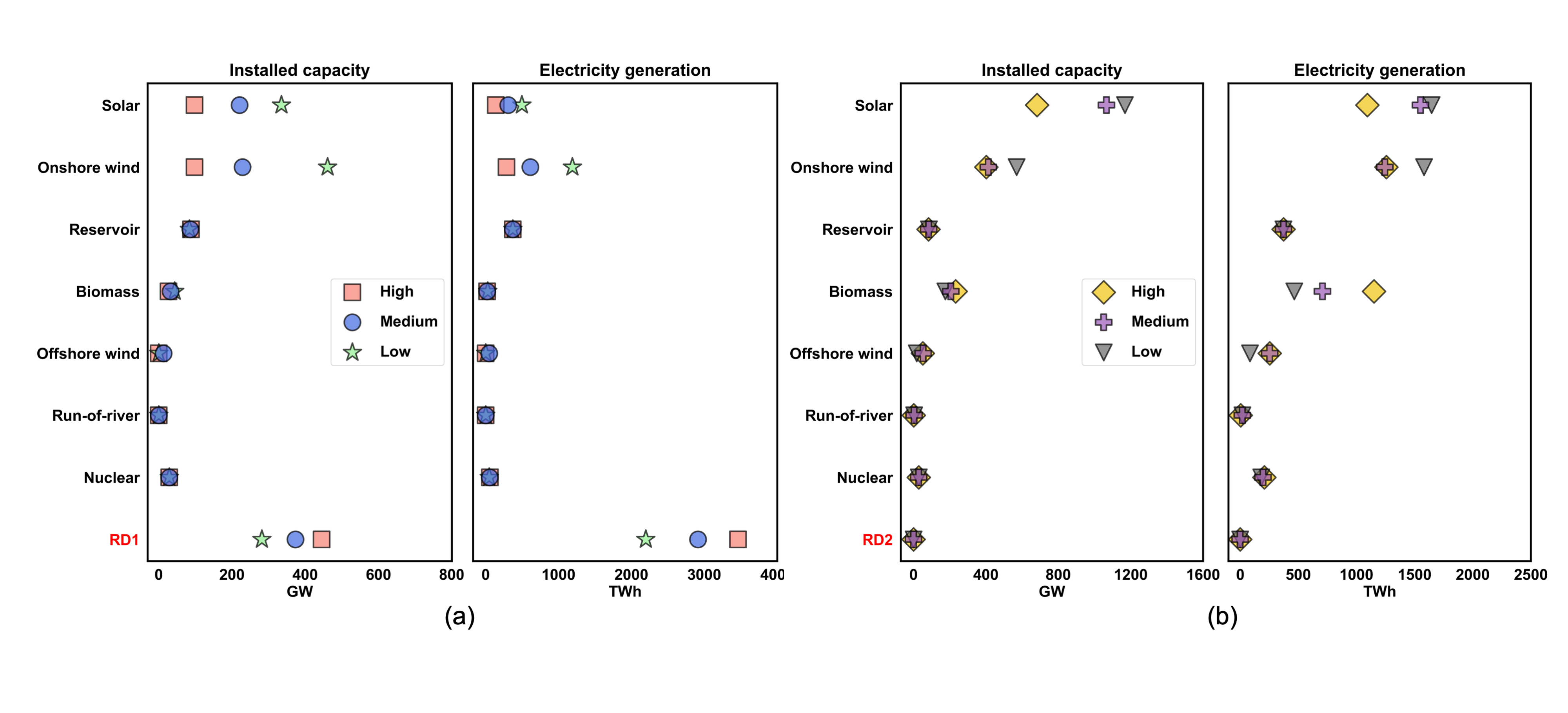} 
    \caption{\textbf{Technology-specific installed capacity and electricity generation across scenarios in 2050.} Panels (a) and (b) show installed capacity and electricity generation by technology under 3 market opportunity scenarios with heliostat design (RD1) and planar design (RD2).}
    \label{fig:2050_results}
\end{figure}

Fig. \ref{fig:2050_results} provides an overview of the installed capacity and electricity generation of various energy sources under different capital cost scenarios for 2050. The capital costs of heliostat design (RD1) and planar design (RD2) are reduced to 6.4–9.7 times and 9.5–14.3 times of solar power in 2050. 

This substantial reduction of costs enhances the competitiveness and viability of heliostat design (RD1) in the 2050 scenarios. With heliostat design, terrestrial wind and solar capacities shrink by up to 75\%, while hydropower remains largely unchanged due to capacity constraints and the assumption of using existing capacity in Europe. 
Due to heliostat design's lower generation costs and higher reliability, the generation from solar and wind decreases by 70-87\% and 29-81\%, respectively. Heliostat design outperforms both solar and wind across all scenarios, emerging as the dominant electricity generation source. 

By contrast, planar design (RD2) never appears in the cost-optimal mix under NASA’s 2050 assumptions, suggesting it is not yet competitive with terrestrial renewables or heliostat design’s more continuous power profile. Collectively, these findings underscore how a substantial reduction in SBSP capital costs can reshape future energy systems, with heliostat design emerging as a potentially dominant baseload resource while planar design lags behind in economic viability under the NASA's current predictions.

\begin{figure}[htbp]
    \centering
    \includegraphics[width=1\textwidth]{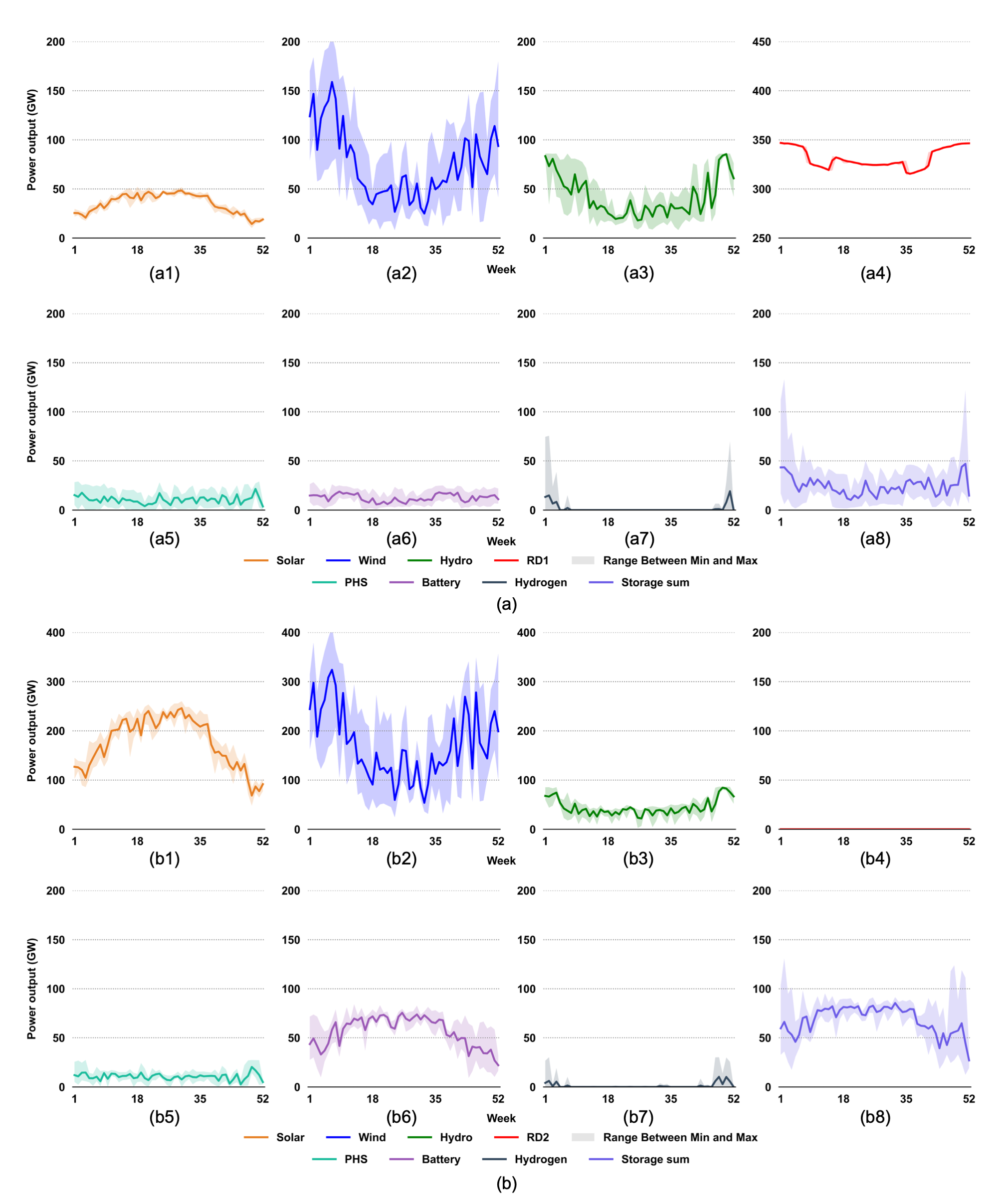} 
    \caption{\textbf{Weekly Power Output Variability of SBSP, Other Renewables and Storage Technologies (Discharging) under Medium SBSP Market Opportunity Scenario in 2050.} Panels (a) and (b) present a comparative analysis of heliostat design (RD1) and planar design (RD2), respectively, with Solar, Wind, Hydro, Pumped Hydro Storage (PHS), Battery, and Hydrogen storage, in terms of power output and variability. Solid lines indicate weekly average power, while shaded bands denote the range of weekly maximum and minimum power fluctuations.}
    \label{fig:generation_storage_2050}
\end{figure}

To assess the impact of SBSP integration, we select the Medium SBSP Market Opportunity Scenario as a representative case. Figure~\ref{fig:generation_storage_2050} compares the weekly average power outputs of SBSP, solar, wind, hydro, and storage technologies under this scenario. Throughout the year, heliostat design consistently delivers 300–350 GW with only minor dips around the spring and autumn equinoxes. By contrast, solar, wind, and hydro in both scenarios hover below 200 GW on average and exhibit notable seasonal swings—solar peaks in summer, while wind and hydro tend to peak in winter or other off-peak solar seasons.

With heliostat design’s inclusion, wind generation declines by approximately 50\%, while solar output drops by 70–80\%, illustrating heliostat design’s displacement of intermittent renewables. In contrast, the planar design (RD2) is not selected as shown in Figure~\ref{fig:generation_storage_2050}b (a no-SBSP case) in which the energy system relies more heavily on variable renewables throughout the year. This increases the need for storage to ensure supply-demand balance. The total storage usage displays larger capacity and wider fluctuations throughout the year than that in the scenario with RD1, underscoring the need for intensive storage interventions in the absence of SBSP.

The deployment of heliostat design leads to a substantial shift in storage utilization. In summer months, average power withdrawn from storage decreases by over 50\% relative to the case without SBSP, primarily due to the consistent and high output from heliostat design reducing the need for balancing services. This is reflected in narrower deviation bands for all storage technologies, indicating a more consistent operating regime. In winter, however, total weekly storage requirements increase slightly, revealing a continued need for long-duration seasonal solutions. Although battery usage declines dramatically (by 78\% annually), hydrogen storage takes on a larger role in winter, highlighting how heliostat design’s near-baseload profile mitigates {short-term} imbalances without fully eliminating {seasonal} challenges.

\begin{figure}[htbp]
    \centering
    \begin{subfigure}[b]{1\textwidth}
        \centering
        \includegraphics[width=0.9\textwidth]{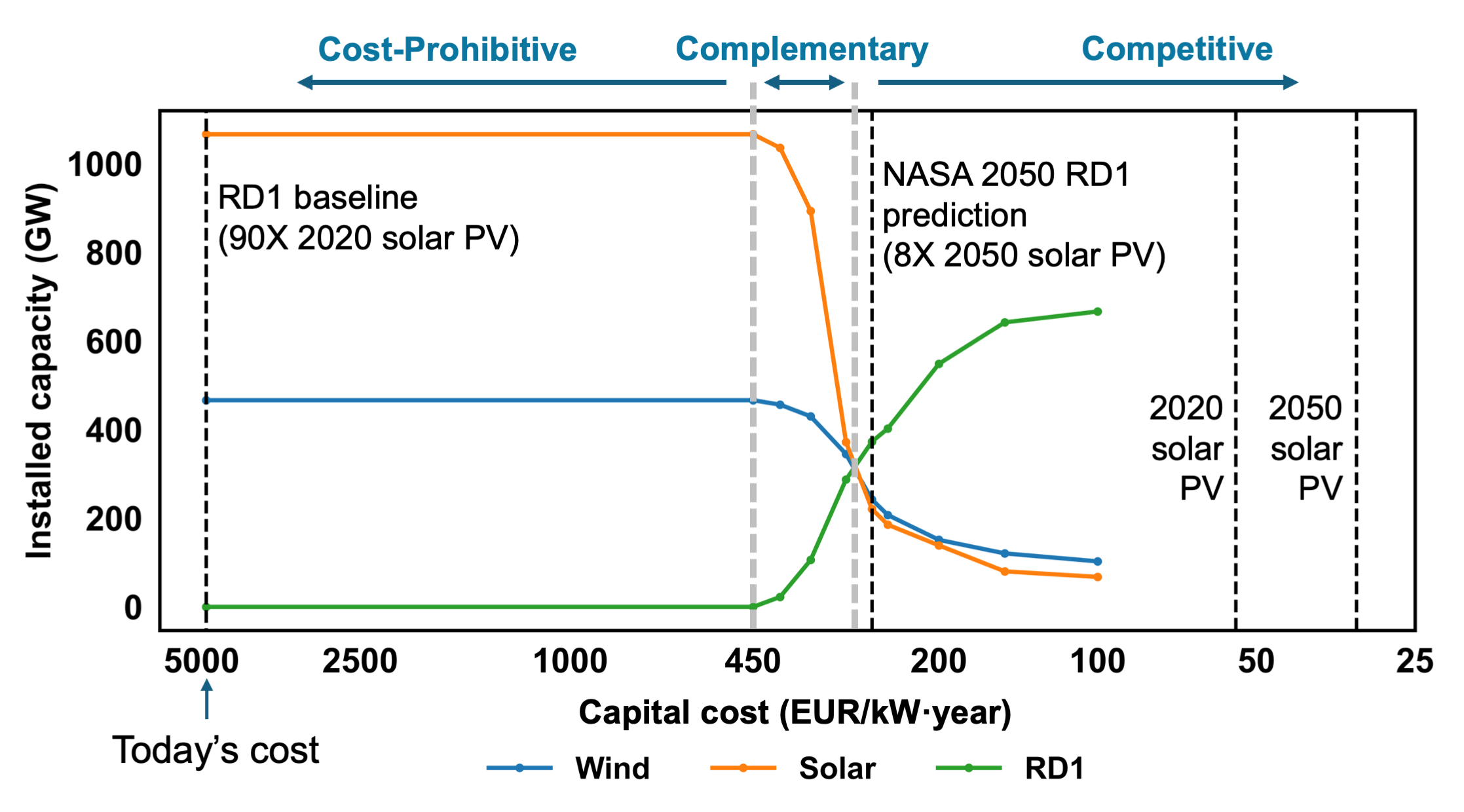}
        \caption{}  
        \label{fig:subfig_a}
    \end{subfigure}
    
    \vspace{1em} 
    
    \begin{subfigure}[b]{1\textwidth}
        \centering
        \includegraphics[width=0.9\textwidth]{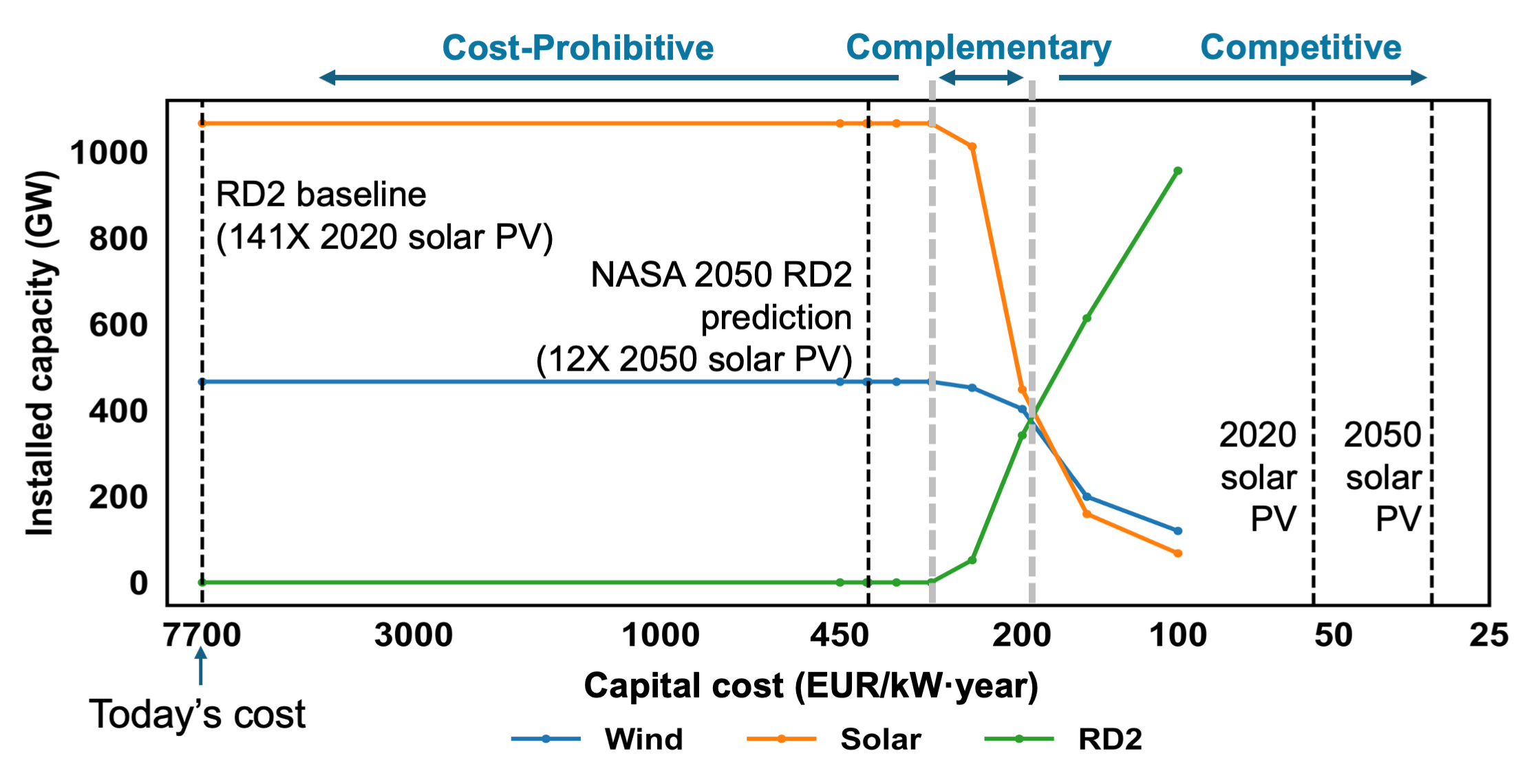}
        \caption{}  
        \label{fig:subfig_b}
    \end{subfigure}
    
    \caption{\textbf{Changes in Installed Capacity in Response to SBSP Capital Costs in 2050.} The panel (a) and (b) compares the installed capacities of terrestrial solar, wind, heliostat design (RD1) and planar design (RD2) under varying SBSP capital costs. The x-axis represents the capital cost of SBSP in EUR/kW$\cdot$year, while the y-axis represents the corresponding installed capacity (GW).}
    \label{fig:combined_sensitivity_capacity}
\end{figure}

\subsection*{Cost thresholds for SBSP in the future (sensitivity analysis)} 

Building on our 2050 scenario findings that rely on NASA’s cost projections, we next conduct a sensitivity analysis to reflect the uncertainty of these estimates and to identify plausible cost thresholds for SBSP compared to terrestrial renewables (i.e., solar). Specifically, we adjust SBSP capital costs by up to ±65\% of heliostat design’s baseline values—100–450 EUR/kW$\cdot$year—representing both optimistic and more pessimistic trajectories. This range ensures the net-zero role exploration that at the upper end, SBSP remains non-competitive, while at the lower end, it can substantially displace other generation options. Throughout these runs, the costs of other technologies remain fixed at their 2050 estimates (``Medium'').


Fig. \ref{fig:combined_sensitivity_capacity} shows that once SBSP capital costs fall to around 14× or 9× the 2050 solar PV cost in heliostat design and planar design cases respectively, SBSP starts displacing wind and solar in a complementary role. A further approximately ~35\% cost reduction beyond these thresholds enables SBSP to emerge as the dominant baseload technology, surpassing conventional renewables in capacity deployment. However, at these levels, SBSP may produce surplus generation (due to reduced load factors, see SI Section 5), effectively meeting both baseload demand and balancing short- and long-duration fluctuations. However, current prohibitive costs of SBSP are at least 1-2 orders of magnitude higher than these complementary or baseload cost benchmarks.  



\begin{figure}[htbp]
    \centering
    \includegraphics[width=1\textwidth]{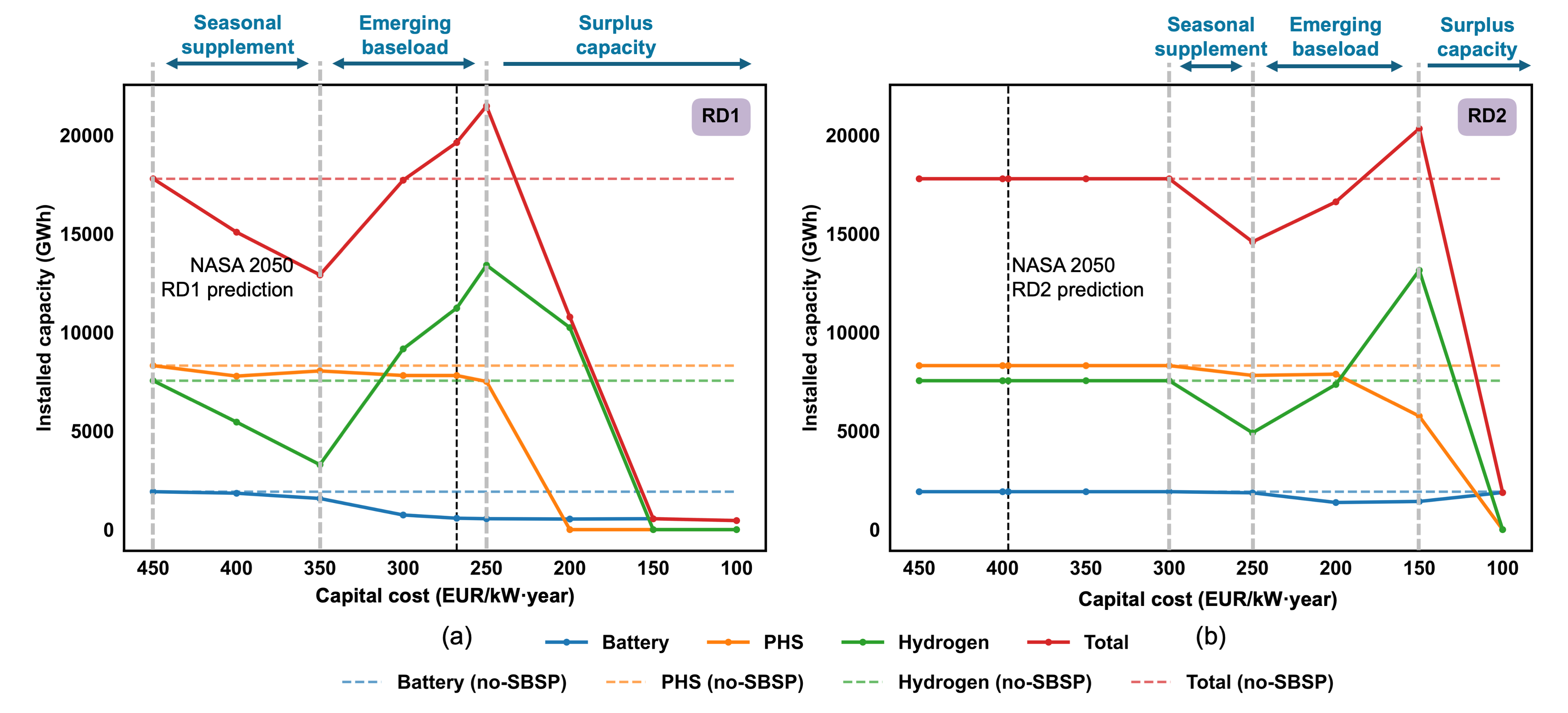} 
    \caption{\textbf{Impact of Heliostat Design (RD1) and Planar Design (RD2) Capital Cost on Optimized Energy Storage Capacity in 2050.} The panel (a) and (b) illustrates the relationship between SBSP’s capital cost and the optimized energy storage capacity of three storage technologies: battery, hydrogen, and pumped hydro storage (PHS), as well as their total capacity, in the 2050 scenario. The energy storage capacity represents the maximum amount of electricity that storage devices can deliver when fully charged.}
    \label{fig:sensitivity_capacity_storage}
\end{figure}

The results further highlight the shifting role of long-duration storage as SBSP scales from small to large deployment (Fig. \ref{fig:sensitivity_capacity_storage}). Initially, as SBSP capital costs decline to approximately 14× (heliostat design) or 9× (planar design) the projected 2050 solar PV cost, the capacity of short- and medium-duration storage technologies, such as batteries and PHS, steadily decreases. Hydrogen storage follows a similar trend, with a sharp reduction occurring when SBSP costs reach around 11–14× (heliostat design) or 8–9× (planar design) the 2050 solar PV benchmark. In this phase, SBSP effectively serves as a seasonal supplement, partially displacing the need for hydrogen storage.

\begin{figure}[htbp]
    \centering
    \includegraphics[width=1\textwidth]{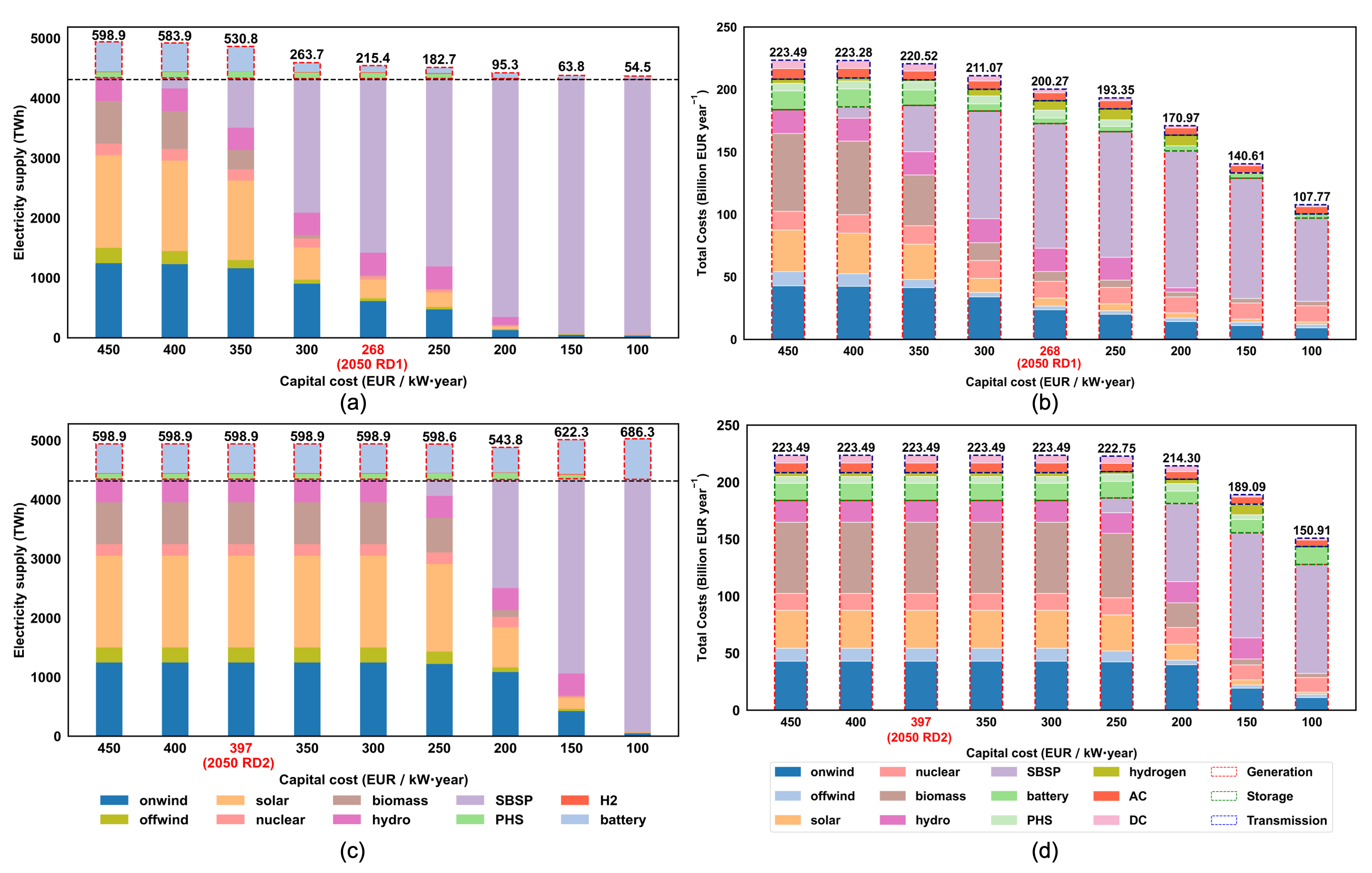} 
    \caption{\textbf{Impact of Heliostat Design (RD1) and Planar Design (RD2) Capital Cost on Annual Electricity Supply and System Costs in 2050.} Panel (a) and (c) show the total annual electricity supply from all generation and storage technologies under varying capital costs of heliostat design and planar design, respectively. Numbers above each bar represent the combined electricity supply from the three storage technologies, and the black dashed line indicates the annual electricity demand. Panel (b) and (d) present the corresponding total annual system costs for each scenario. Numbers above each bar indicate the total annual costs under different SBSP capital cost assumptions.}
    \label{fig:sensitivity_cost_generation}
\end{figure}

However, once SBSP capital costs fall below approximately 11× (heliostat design) or 8× (planar design) the 2050 solar PV cost, hydrogen storage capacity surges dramatically, reaching up to 178\% (heliostat design) and 174\% (planar design) of its original level in the absence of SBSP. This sudden expansion indicates that at these cost levels, winter electricity demand imbalances drive extensive reliance on hydrogen storage, despite SBSP’s ability to manage short-term fluctuations. Indeed, while total storage capacity increases during this phase, the actual energy supply from storage continues to decline (Fig.\ref{fig:sensitivity_cost_generation}). In this context, hydrogen provides up to 96\% of its total annual supply during winter, highlighting the persistent need for long-duration storage to complement SBSP, even when SBSP reaches cost-competitive levels.

A sharp decline in total storage capacity occurs when SBSP capital costs drop further to 8× (heliostat design) or 5× (planar design) the 2050 solar PV cost, marking the point at which SBSP surplus capacity becomes dominant. This shift signals SBSP’s ability to provide continuous baseload power, significantly reducing the system’s reliance on both short- and long-duration storage solutions.

Figure \ref{fig:sensitivity_cost_generation}a and \ref{fig:sensitivity_cost_generation}c illustrate the transition of SBSP into the dominant electricity source as its capital cost decreases. When SBSP capital costs reach approximately 9× (heliostat design) or 6× (planar design) the projected 2050 cost of solar PV, SBSP surpasses other technologies in electricity generation. A further decline to around 3× the solar PV cost results in heliostat design and planar design contributing 99\% and 98\% of total electricity generation, respectively.

Notably, heliostat design and planar design exhibit opposing impacts on energy storage requirements. As heliostat design deployment increases, total storage supply declines, shrinking to just 9\% of its initial level. In contrast, planar design—due to its greater output variability—necessitates additional storage to compensate for fluctuations. Despite these divergent trends in overall storage supply, the share of battery consistently increases across both scenarios. When SBSP capital costs fall to 3× the cost of solar PV, electricity supply from battery accounts for 100\% of total stored energy in both heliostat design and planar design cases. This finding underscores a critical threshold: if SBSP costs decline sufficiently, a power system without long-term storage could become viable.

Fig. \ref{fig:sensitivity_cost_generation}b and \ref{fig:sensitivity_cost_generation}d show that the total annual system cost declines significantly as SBSP capital costs decrease, once heliostat design or planar design is deployed. When their capital costs fall to 3× the projected 2050 solar PV cost, total system costs are reduced by 52\% for heliostat design and 33\% for planar design, underscoring SBSP’s substantial potential to lower overall energy expenditures. Throughout this transition, SBSP’s total generation cost initially rises due to increased deployment but subsequently declines sharply as generation costs decrease, ultimately displacing biomass and wind to become the dominant contributor to generation costs. Additionally, the reduction in SBSP generation costs leads to a corresponding decline in transmission costs, with DC transmission experiencing the most pronounced decrease. This finding further supports the conclusion from Fig. S8, which highlights SBSP’s superior ability to reduce long-distance DC transmission requirements.

\section{Discussions}

This study explores the system-wide impacts of two representative SBSP designs—heliostat design (RD1, near-constant, lower TRL) and planar design (RD2, partially intermittent, higher TRL)—under diverse European energy system conditions. When we apply NASA’s 2050 cost and performance assumptions, heliostat design consistently emerges as a cost-competitive option, capable of reducing total system costs by 7–15\% relative to a no-SBSP scenario. Its near-baseload profile not only displaces a substantial share of wind and solar but also cuts battery storage usage by as much as 78\%. At the same time, long-duration solutions, especially hydrogen, still expand in winter months, confirming SBSP’s ability to resolve short-term imbalances while underscoring ongoing seasonal needs. In contrast, planar design never appears in the least-cost mix at NASA’s 2050 cost projections, implying that its improved but still partially intermittent orbital geometry and higher capital expenses render it uncompetitive against both heliostat design and terrestrial renewables. Notably, the stable, cross-network generation offered by heliostat design’s power supply markedly reduces long-distance DC transmission by 31–70\% across regions, suggesting that a centrally coordinated, high-availability resource can diminish reliance on interregional electricity flows.

Our sensitivity analysis pinpoints the capital-cost thresholds at which each design moves from non-competitive to supplementary, and ultimately to a dominant baseload resource. At present, SBSP’s costs are 1–2 orders of magnitude above these break-even points. As heliostat design and planar design costs approach 9–14× that of 2050 solar PV, SBSP begins to complement terrestrial renewables; dropping further to around 6–9× enables a baseload role with surplus capacity for balancing. Below these thresholds, SBSP could start to become the primary generation source, though the system might still reply on energy storage. Conversely, planar design—given its partially intermittent orbit—would also require cost reductions to at least match the lower complementary threshold, plus a reliance on short-duration storage to handle variability.

Collectively, our findings show that near-constant SBSP (as exemplified by heliostat design) may deliver significant system benefits, but also depends on technological breakthroughs (e.g., advanced orbital assembly) to overcome its currently low TRL. In the shorter term, less-intermittent planar design, with higher TRL, may make it an easier candidate for early demonstration, but only if its capital costs are reduced sufficiently to remain economically viable alongside terrestrial solar, wind, and energy storage. A coordinated development strategy could thus focus on planar design demonstrations first to refine core SBSP technologies—such as wireless power transmission and modular in-orbit assembly—while concurrently accelerating R\&D for SBSP designs with more continuous power generation. Such a plan balances the practical realization of a tangible full-scale SBSP prototype with the longer-term potential for near-baseload SBSP to disrupt the broader renewables landscape once its more complex capabilities achieve higher maturity.

To achieve the cost-competitiveness of SBSP, achieving a gigawatt-scale SBSP platform by the 2030s requires several key breakthrough advances. Although space-qualified photovoltaics can achieve high efficiency, deploying vast arrays—potentially on the order of several or tens of km² in a single platform—remains unproven. In-orbit manufacturing and autonomous assembly, both at low to mid TRL, need to advance in parallel with large-scale orbital robotics and modular construction. Equally crucial is wireless power transmission. While small-scale demonstrations, such as Caltech’s recent low-power orbit-to-Earth test \cite{ayling2024wireless}, validate core concepts, maintaining and scaled-up operating highly focused microwave beams over tens of thousands of kilometers needs to be proven. Consequently, our capacity-expansion modeling, though primarily focused on the system-level economic and dispatch implications of SBSP, must treat the 2030 horizon as an early-stage full-scale demonstration phase, with true commercialization more plausibly materializing before 2050 if all required technical advancements progress in parallel.

The concurrent rise of commercial spaceflight and mass satellite manufacturing provides an opportunistic industrial backdrop for SBSP. Companies like SpaceX and Blue Origin are drastically lowering launch costs, while mega-constellations demonstrate high-volume spacecraft production. Such synergies could substantially reduce the marginal expenses of placing SBSP hardware in orbit, offsetting one of the biggest historical barriers. Our model-based findings, which incorporate optimistic launch cost forecasts, highlight how SBSP might compete with terrestrial renewables if launch rates and orbital assembly processes advance quickly. Beyond pure economics, SBSP could also improve European energy security if it helps reduce dependency on imported fossil fuels—aligning with the policy imperative for diversified, domestically controllable generation.

While this study provides valuable insights into SBSP integration within Europe’s energy system, several limitations should be acknowledged. First, the modeling of SBSP simplifies orbital energy dynamics by assuming a fixed shadow period and disregarding transient power fluctuations during Earth’s eclipse phases. Additionally, the study restricts SBSP deployment to equatorial GEO at 0° longitude, omitting considerations of alternative orbital placements or variations in altitude that could impact energy yield and transmission efficiency. Second, the modeling does not account for potential impacts from space-specific challenges such as orbital congestion, transmission interruptions, or beaming variability, which could influence SBSP reliability and operational performance. Further work will take these factors into consideration for improving the generation characteristics in the model. Third, the construction of the 2050 energy system relies on 2020 climate data due to the absence of future climate data, which may not fully capture long-term meteorological trends affecting renewable generation. Fourth, the analysis does not assess SBSP’s role in grid stability, particularly its ability to provide ancillary services, manage fluctuations, or complement terrestrial renewables in real-time operations—an essential factor for large-scale integration. Fifth, the study does not account for geopolitical, regulatory, or spatial constraints that may impact SBSP deployment across different European nations, potentially limiting its feasibility in real-world implementation. Sixth, our SBSP performance and costs derive from NASA’s reports, and may not encompass the full spectrum of perspectives in the broader SBSP community, although our sensitivity analysis aims to mitigate this limitation. In particular, advanced space power measurement techniques, such as interval-oriented eigensystem realization algorithms \cite{YANG2025118929}, reliability-constrained sliding mode control approaches \cite{10843835}, and multi-objective optimization-inspired regularization methods \cite{YANG2025117814}, are emerging as critical tools to better characterize SBSP performance in orbit. 

Looking ahead, future SBSP research should prioritize the coordinated advancement of enabling technologies such as multi-junction and lightweight photovoltaic cells, modular in-orbit assembly methods, and wireless power transmission techniques, while also fostering close collaboration with policymakers to ensure alignment with emerging regulatory frameworks. Future studies should also focus on developing strategies to mitigate the risks posed by orbital debris to SBSP systems \cite{CHOI202427}. In addition, enhancing the resilience of SBSP system architectures and strengthening encryption protocols for secure energy transmission will be essential to protect against potential hostile actions. Improvements in module performance and cost-effectiveness, together with proactive policy engagement, will be critical to move SBSP from conceptual feasibility toward scalable implementation.

At the system level, future research should expand beyond Europe to explore global SBSP deployment opportunities, incorporating more precise satellite positioning, ground receiver optimization, and real-time operational feasibility, including the provision of ancillary services. Future work should also explore the role of SBSP in supporting a more integrated energy system by incorporating cross-sectoral electrification and sector coupling, enabling a better assessment of SBSP’s contribution to system-wide flexibility and decarbonization. Additional efforts are needed to optimize hybrid configurations integrating multiple SBSP designs, assess the feasibility of intercontinental SBSP networks for regional balancing, and investigate social acceptance, electromagnetic safety, and regulatory challenges. Evaluating risks associated with heavy reliance on orbital infrastructure, such as space interference, system degradation, and equipment failure, alongside developing collaborative international frameworks, will be essential to ensure resilient, equitable, and sustainable energy transitions.

Beyond SBSP, some other emerging technologies promise firm, low-carbon power or help mitigate renewables’ variability. Advanced nuclear—encompassing small modular reactors (SMRs) and Gen~IV designs, with fusion more distant—aims to deliver continuous, carbon-free electricity. Demonstration SMRs have entered operation or advanced construction in recent years \cite{IAEA2021}, and a Gen~IV pebble-bed reactor began supplying power in 2021 \cite{WNN2021HTRPM}. By contrast, SBSP is unlikely to mature before the 2040s, so advanced nuclear may achieve commercial readiness sooner.

For flexibility, low-carbon fuels and storage technologies can balance intermittent renewables over diverse timescales. A recent European pilot successfully ran a 12 MW turbine on 100\% hydrogen in 2023 \cite{IEA2024Hydrogen}, illustrating the feasibility of hydrogen-based peaking or seasonal supply. Meanwhile, other long-duration storage solutions (flow batteries, compressed air, pumped hydro) can buffer multi-day shortfalls in wind or solar \cite{HE2021100060}.

Taken together, these technologies offer a complementary portfolio. SBSP and advanced nuclear provide baseload-like generation, reducing storage needs; excluding firm low-carbon resources can drive steep cost increases near 100\% renewables \cite{SEPULVEDA20182403}. Although SBSP (low TRL, high uncertainty) and advanced nuclear (moderate TRL, complex licensing) each face distinct obstacles, integrating multiple firm and flexible options may yield a more resilient, cost-effective pathway to net zero in Europe. More research on the trade-off and uncertainty between these technologies is needed for insights of cost-effective, reliable net zero transition.

\section{Methods}
\begin{figure}[htbp]  
    \centering
    \includegraphics[width=\textwidth]{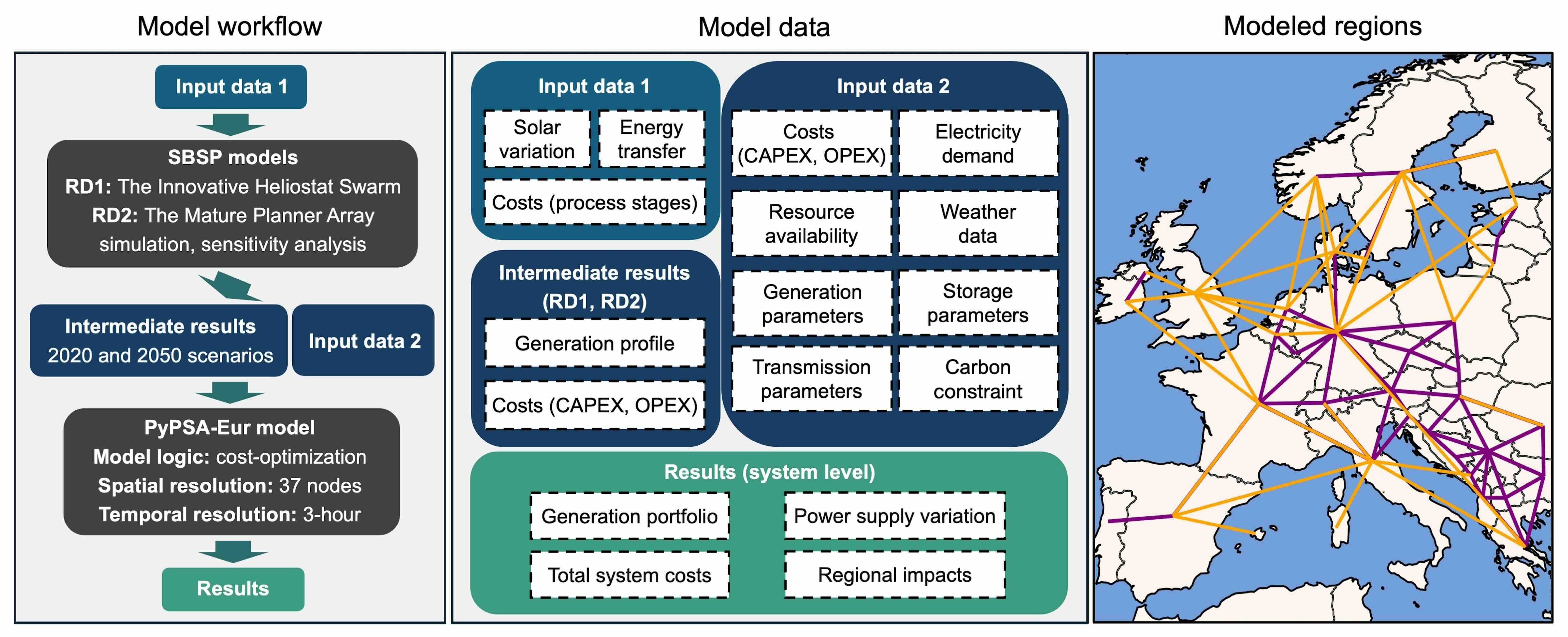}  
    \caption{\textbf{Overview of model workflow, data and regions.} The structure of this workflow diagram is inspired by the approach outlined in \cite{sasse2020regional}, while the content has been independently developed to represent the specific processes and parameters of this study.}
    \label{fig:workflow}  
\end{figure}

\noindent \textbf{Modelling approach}
Our modelling approach (Fig. \ref{fig:workflow}) soft links the SBSP model and the PyPSA-Eur \cite{PyPSAEur} model to assess the cost-competitiveness of SBSP for European power system decarbonization in 2020 and 2050. Initially, we model the power generation of various SBSP designs using the methodology outlined in Section 2, incorporating generation profiles and cost data for heliostat design and planar design. Next, we run the PyPSA-Eur model to optimize power generation scenarios for 2020 and 2050, performing time-step optimizations in 3-hour intervals. This includes validating the 2020 network as a reference for further sensitivity analyses. Finally, different SBSP designs are integrated into the PyPSA-Eur model as generators, and re-optimized to assess the impact of SBSP on both the current and future European electrical systems, as well as its technical feasibility. Across various scenarios, we analyze system generation costs, total energy generation by different sources, the relationship between major energy sources and SBSP costs, greenhouse gas emissions, and the countries benefiting from SBSP deployment.

\noindent \textbf{PyPSA-Eur model}
The model is a freely accessible dataset that represents the European energy system at the transmission grid level, covering the entire European Network of Transmission System Operators for Electricity (ENTSO-E) region \cite{PyPSAEur}. While the purpose of the SBSP model is to generate cost estimates and power generation profiles for SBSP designs, the role of PyPSA-Eur is to simulate the European electricity system across different years, offering a platform for computations of electricity generation and the assessment of storage and transmission requirements. The PyPSA-Eur model incorporates 37 nodes representing a simplified version of the grid for our study region. The primary objective of PyPSA-Eur is to minimize the total annualized system costs for each scenario. To manage the computational complexity of running multiple scenarios, we employ k-means \cite{PyPSA} clustering to simplify the grid layout and perform optimizations at 3-hour intervals.


\noindent \textbf{Networks and Countries}
As outlined in \cite{PyPSAEur}, the majority of countries in our model are represented by a single node. Exceptions include four countries—the United Kingdom (GB), Spain (ES), Italy (IT), and Denmark (DK)—which are each modeled with two nodes. In total, the model spans 33 countries, consisting of the 28 European Union member states as of 2018 (excluding Cyprus and Malta), in addition to Bosnia and Herzegovina (BA), Norway (NO), Serbia (RS), and Switzerland (CH). These countries encompass the primary synchronous zones of ENTSO-E. The nodes are interconnected through a network built on both existing and planned transmission line links between the countries. Specifically, the network also incorporates a comprehensive range of technologies for both renewable and conventional electricity generation, along with energy storage and transmission systems. Renewable energy sources include onshore and offshore wind (with floating, AC, and DC variants), solar PV (both standard and high solar altitude tracking configurations), and hydropower from run-of-river, biomass, and geothermal sources. Conventional power generation is provided by nuclear, coal (hard and lignite), gas (using both combined-cycle and open-cycle gas turbines), and oil. Energy storage options feature battery systems, hydrogen storage, reservoir \& dam systems, and pumped hydropower storage. Transmission infrastructure comprises high voltage alternating current (HVAC) and high voltage direct current (HVDC) lines. Additionally, the structure of the network, including the countries, nodes, types of electricity generation, energy storage options, and transmission systems, remains largely unchanged between the 2020 and 2050 scenarios.

\noindent \textbf{Electricity Demand}
In this study, electricity demand refers exclusively to electricity consumption without including the electrification of transport, heating, or other emerging sectors, both in the 2020 and 2050 scenarios. The hourly electricity demand profiles for each country in 2020 are sourced from the ENTSO-E website \cite{entsoe_load_data}. For 2050, we adopt the “Native Demand” projections from the ENTSO-E \& ENTSOG TYNDP 2024 Scenarios \cite{TYNDP2024} (Global Ambition, climate year 2009), which represent electricity demand from centralized market, residential, and tertiary sectors, while explicitly excluding additional demands associated with battery charging, electric vehicles (EVs), heat pumps, electrolyzers, and other sector-coupled loads.

\noindent \textbf{Electricity Supply and Storage}
For both the 2020 and 2050 scenarios, data for power plants is sourced using the open-access powerplantmatching tool \cite{gotzens_performing_2019}, which compiles datasets from multiple sources. The tool provides detailed information on plant location, technology type, fuel source, age, and capacity. It includes various generation types such as coal, lignite, oil, open-cycle and combined-cycle gas turbines (OCGT and CCGT), nuclear power, geothermal energy, bioenergy, and hydropower. For the 2050 scenario specifically, we remove power plants with a DateOut extending beyond 2050 to ensure data consistency.

Renewable generation options that can be expanded include onshore and offshore wind, solar photovoltaics, concentrated solar power, and run-of-river hydropower. For both 2020 and 2050, generation time series and potential are calculated based on 2020 ERA5 weather data \cite{https://doi.org/10.1002/qj.3803}. Land eligibility restrictions, such as suitable land classifications and buffer zones around populated or protected regions, are applied to wind and solar power to determine feasible capacity.

Additionally, in our model, the storage module includes batteries, existing pumped-hydro storage (PHS), hydroelectric dams, hydrogen storage, and synthetic energy carriers such as methane. It is worth noting that although hydrogen has potential applications across multiple sectors, in this study, hydrogen storage is modeled solely for electricity balancing purposes, as our analysis focuses exclusively on the electricity sector.

\noindent \textbf{Technology and cost assumptions}
For the technology and cost assumptions, we use 2020 data for the current network and projected 2050 values for various technologies, including parameters such as investment costs, fixed and variable operation and maintenance (FOM and VOM) costs, lifetimes, and efficiencies. Many of these figures are sourced or revised from the technology database published by the DEA \cite{DEA2020} , LAZARD \cite{Lazard2023LCOE} and the IEA’s Global Energy and Climate Model (GEC) \cite{IEAGlobal2023}. 
The overnight capital costs are converted to net present costs by applying an annualisation factor, which uses a discount rate \(r\) over the asset’s economic lifetime \(n\):
\begin{equation}
a = \frac{1 - (1 + r)^{-n}}{r},
\end{equation}
From these parameters, the \textit{marginal\_cost} and \textit{capital\_cost} in models are then computed automatically \cite{PyPSAEur}.
Furthermore, to ensure a more comprehensive and robust assessment, we incorporate three sets of cost data of 2050—low, mid, and high values—when inputting costs for key energy sources into the model. These cost ranges are based on projected price estimates for 2050, allowing us to account for the inherent uncertainty in future cost developments. 
Assumptions are maintained at \href{https://github.com/pypsa/technology-data}{github.com/pypsa/technology-data}, and were taken from version 0.9.2 \cite{PyPSA2024}.

\noindent \textbf{Carbon dioxide emissions}
In our model, we use the \textit{CO$_2$ intensity} provided in \cite{PyPSA2024} for carbon dioxide emissions. The specific intensities are as follows: oil at 0.2571 tCO$_2$/MWh\_th, coal at 0.3361 tCO$_2$/MWh\_th, lignite at 0.4069 tCO$_2$/MWh\_th, gas at 0.1980 tCO$_2$/MWh\_th, and geothermal at 0.1200 tCO$_2$/MWh\_th. All other energy sources are assumed to have zero CO$_2$ emissions. Since the launch of the European Union’s greenhouse gas (GHG) reduction strategy in 2005, emissions from the power and heat generation sectors, as well as energy-intensive industries covered by the European Emissions Trading System (EU ETS), have decreased by approximately 43\%. Complementary regulations, such as those promoting renewable energy and energy efficiency, have played a significant role in helping the EU surpass its target of reducing GHG emissions by 20\% by 2020 compared to 1990 levels. In fact, by 2020, the EU had achieved a reduction of around 31\% relative to 1990 \cite{EU2021carbon}. In line with this, we incorporated a \textit{GlobalConstraint} into our 2020 model, setting \textit{co$_2$\_emissions} at roughly 70\% of 1990 levels. Additionally, in accordance with the European Climate Law \cite{EUClimateLaw2021}, which mandates achieving net-zero greenhouse gas emissions across the EU by 2050, we set the 2050 \textit{co$_2$\_emissions} to zero in our model.

\noindent \textbf{Validation}
The purpose of this section is to evaluate the effectiveness of our model by assessing the quality of data for the 2020 scenario. This is achieved by comparing model outputs with publicly available datasets at both continental and country levels across Europe.

We follow the same methodology used in previous PyPSA-Eur studies when modelling the network structure \cite{PyPSAEur, validation1}. These studies have already conducted a detailed validation of both the total line lengths and the network topology. The findings indicate that, when considering data for all countries, the deviation between the line lengths in the PyPSA-Eur dataset and the circuit lengths reported by ENTSO-E shows a mean absolute error of less than 15\% across all voltage levels. Additionally, in terms of network topology, PyPSA-Eur exhibits strong alignment with real-world data, demonstrating good consistency across various regions \cite{PyPSAEur}.

Since we update the electricity demand, generation costs for various energy sources, and power plant data to reflect conditions in 2020, we also model renewable energy potentials, such as solar and wind, based on the ERA5 dataset \cite{Hofmann2021}. The model validation is performed by comparing the installed generation capacities and the outputs from a Linear Optimal Power Flow (LOPF) simulation against historical data, ensuring consistency between the simulated and actual generation performance. 

\begin{figure}[htbp]
    \centering
    \begin{subfigure}[t]{0.49\textwidth}
        \centering
        \raisebox{0.5\height}{\textbf{(a)}}\hspace{-0.5cm}
        \includegraphics[width=\textwidth]{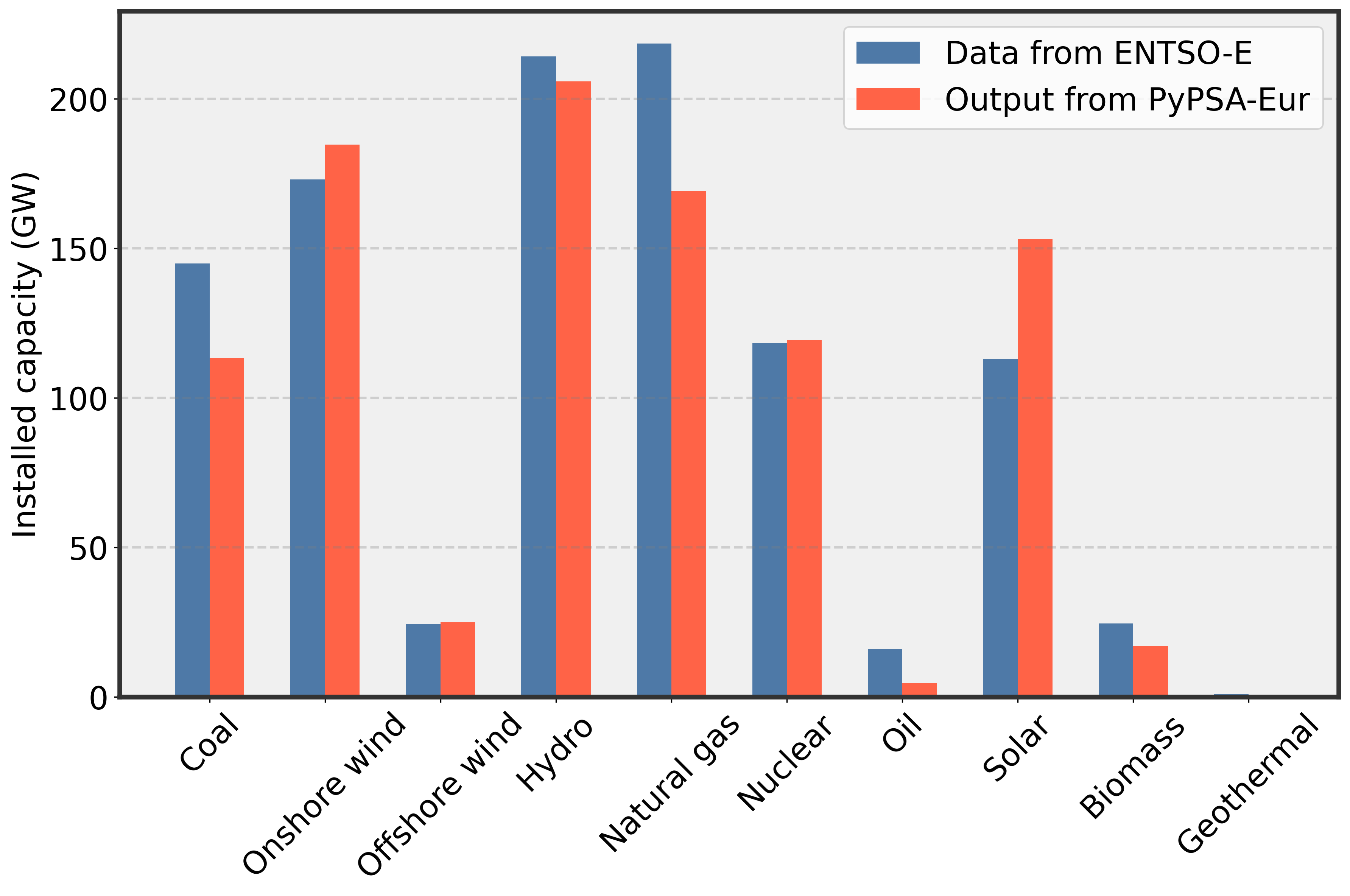}
        \caption{}
        \label{fig:installed_total}
    \end{subfigure}
    \hfill
    \begin{subfigure}[t]{0.49\textwidth}
        \centering
        \raisebox{0.5\height}{\textbf{(b)}}\hspace{-0.5cm}
        \includegraphics[width=\textwidth]{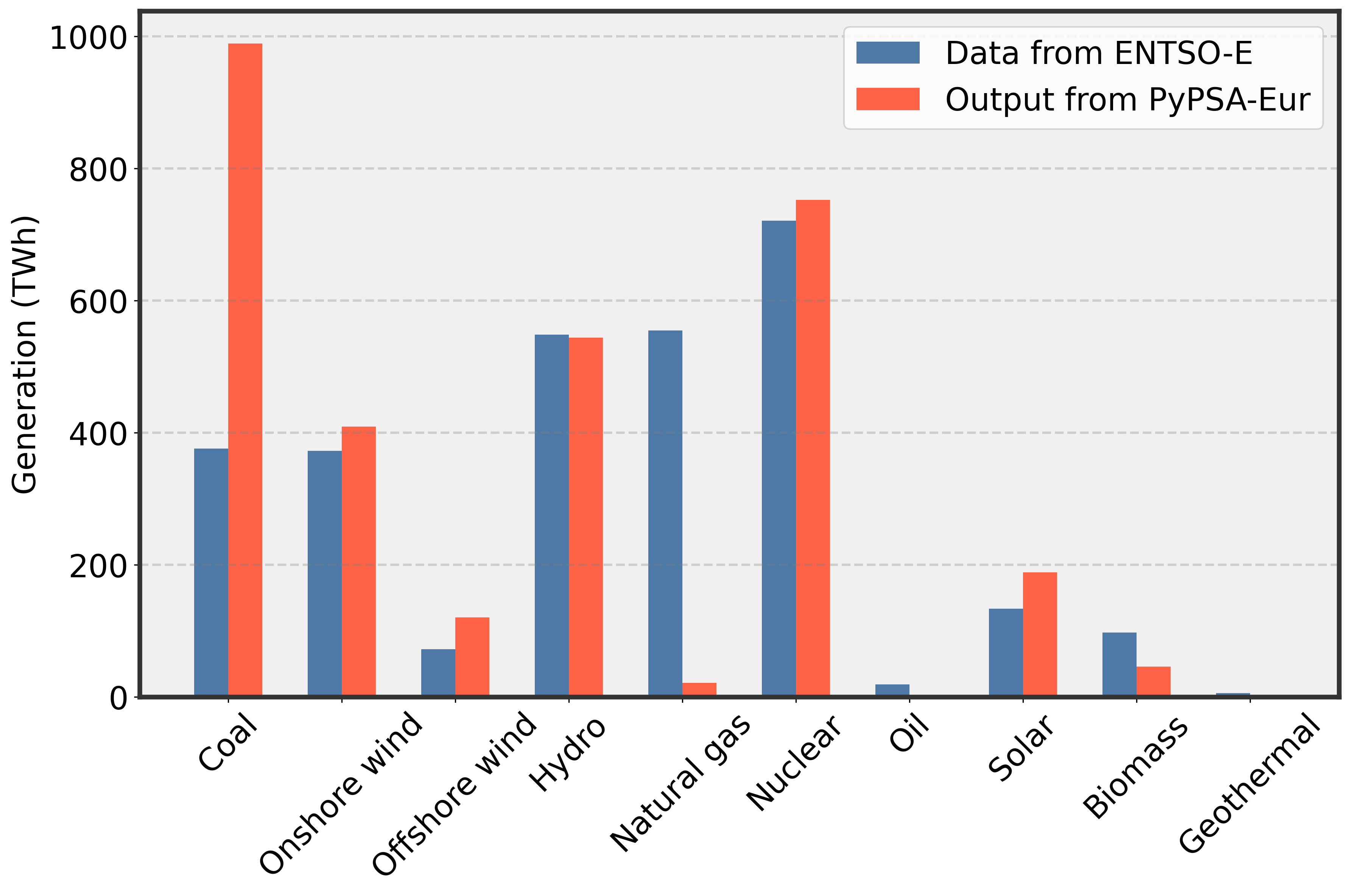}
        \caption{}
        \label{fig:generation_total}
    \end{subfigure}

    \caption{\textbf{Comparison of installed capacity and LOPF simulation of 2020 to historical data published by ENTSO-E.} }
    \label{fig:validation_1}
\end{figure}

\begin{figure}[htbp]
    \centering
    \includegraphics[width=0.9\textwidth]{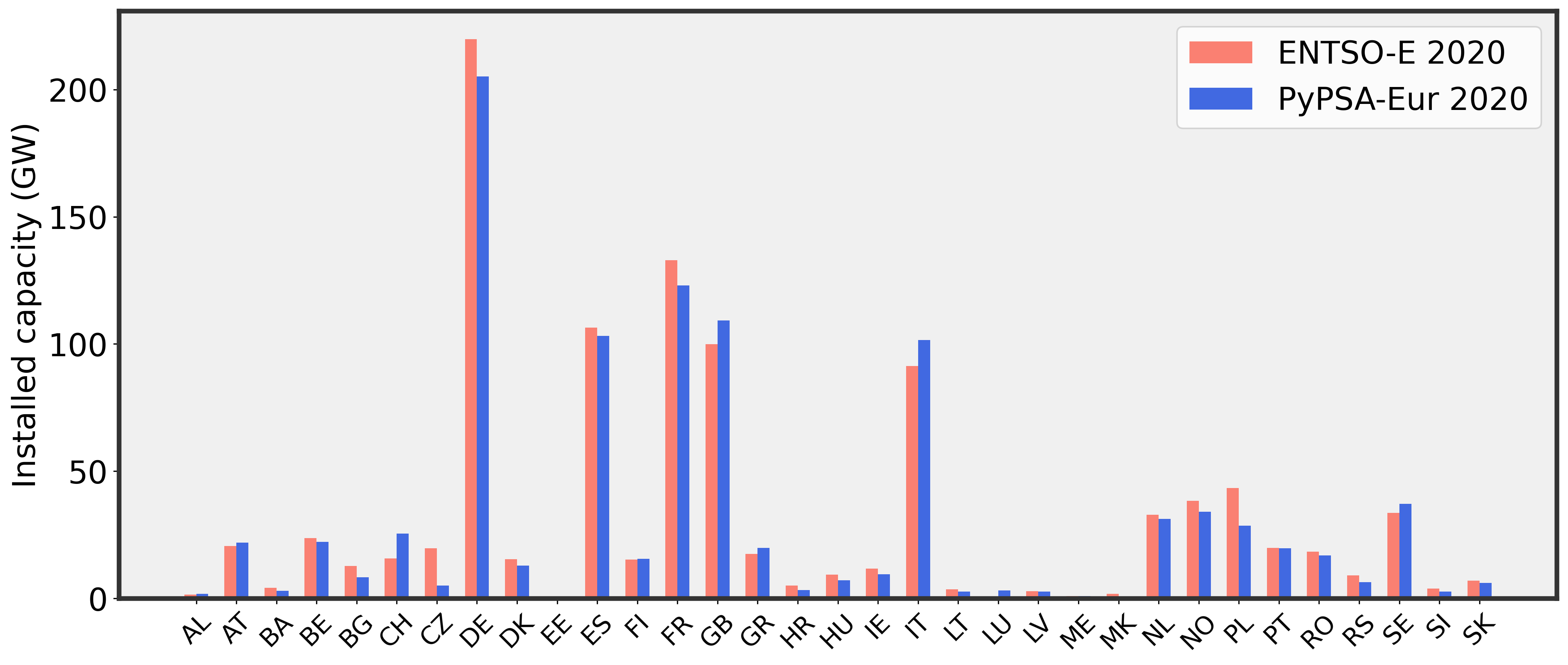} 
    \caption{\textbf{Comparison of installed generation capacity in the EU by country to historical data reported by ENTSO-E}}
    \label{fig:validation_2}
\end{figure}

Figure \ref{fig:installed_total} demonstrates that our model aligns closely with existing databases, capturing 993 GW out of the 1048 GW of installed capacity reported by ENTSO-E \cite{entsoe_actual_generation, entsoe_generation_capacity}. Discrepancies in coal, natural gas, oil, and biomass capacities can largely be attributed to the lag in data updates, which omits recently commissioned plants and results in an underestimation of these capacities in our model. Additionally, using ERA5 data to model solar potential introduces a known bias, as ERA5-derived solar capacity tends to be overestimated \cite{era5}, leading to an inflated solar capacity representation. However, given the model’s inclusion of 33 countries, validation from a technology-only perspective is insufficient to fully assess its accuracy. Figure \ref{fig:validation_2} therefore provides additional validation at the country level, showing that most countries align with reported data to a high degree of accuracy (2\%–15\% error).
Apart from installed capacity, Figure \ref{fig:generation_total} shows that most generation types have comparable power outputs (see Table \ref{tab:validation_1} for numerical data). Beyond minor deviations due to capacity differences, the main variation appears in the generation mix, particularly between coal and natural gas. Higher input marginal prices of natural gas compared to coal, combined with the absence of a CO$_2$ pricing mechanism, result in lower natural gas generation. Looking ahead to 2050 scenarios, as carbon emissions are set to zero, fossil-based sources such as coal, natural gas, and oil will be phased out, rendering this discrepancy non-influential for future analyses.

The validation shows promising results, indicating that it generates the anticipated output, but it is not designed to serve as a fully validated model, due to its reliance on various user-defined inputs, especially for 2050 scenarios. This issue becomes even more pronounced when modeling future energy scenarios, which involve numerous assumptions. Given the range of uncertainties, it is not feasible to create a fully validated model of these future scenarios. Instead, such models should be viewed as tools to deepen our understanding of how energy systems might operate, rather than attempting to perfectly replicate future systems.

\begin{table}[h!]
\centering
\renewcommand{\arraystretch}{1.5} 
\setlength{\tabcolsep}{4pt} 
\caption{Comparison between the generation outputs from the LOPF simulation to historical data.}
\begin{tabular}{p{2.5cm}ccccccc} 
\toprule
 & \textbf{Simulated} & \textbf{Historical} & \textbf{Difference} & \textbf{Simulated (\%)} & \textbf{Historical (\%)} & \textbf{Difference (\%)}\\
\midrule
\textbf{Coal} & 988.7 & 376.2 & 612.5 & 32.2 & 13.0 & 19.2 \\
\textbf{Onshore wind} & 408.9 & 372.6 & 36.3 & 13.3 & 12.8 & 0.5 \\
\textbf{Offshore wind} & 120.7 & 72.2 & 48.5 & 3.9 & 2.5 & 1.4 \\
\textbf{Hydro} & 543.6 & 548.1 & -4.5 & 17.7 & 18.9 & -1.2 \\
\textbf{Natural gas} & 21.6 & 554.6 & -533 & 0.7 & 19.1 & -18.4 \\
\textbf{Nuclear} & 752.5 & 721.0 & 31.5 & 24.5 & 24.9 & -0.4 \\
\textbf{Oil} & 0 & 18.9 & -18.9 & 0 & 0.7 & -0.7 \\
\textbf{Solar} & 188.9 & 133.6 & 55.3 & 6.1 & 4.6 & 1.5 \\
\textbf{Biomass} & 46.1 & 97.3 & -51.2 & 1.5 & 3.4 & -1.9 \\
\textbf{Geothermal} & 0.6 & 5.9 & -5.3 & 0 & 0.2 & -0.2 \\
\textbf{Total} & 3071.6 & 2900.5 & 171.1 & 99.9 & 100 & -0.1 \\

\bottomrule
\end{tabular}
\label{tab:validation_1}
\end{table}

\noindent \textbf{Soft-linked SBSP-PyPSA-Eur model}
Among the two integrated models, the SBSP model emphasizes simulated parameters for two representative designs, heliostat design and planar design, focusing on annual generation profiles and associated costs projected for 2020 and 2050. The PyPSA-Eur model, on the other hand, serves as an electricity system optimization tool that aims to minimize total annualized system costs across various scenarios.

Using heliostat design as an example, we incorporate SBSP as a new generator type into the updated 2020 and 2050 networks. Given that a single SBSP satellite can deliver energy to multiple regions simultaneously \cite{ref1}, we assume each node has a dedicated ground station to receive this energy. Thus, an SBSP generator is added at each node, simulating the simultaneous energy reception across nodes. 
We set the capacity parameter \textit{p\_nom} as \textit{extendable}, allowing it to adjust according to demand. Additionally, \textit{p\_max\_pu} is updated to align with the SBSP generation profile data, normalized for consistency. This modeling approach reflects the treatment of SBSP in a similar way to other variable renewable sources such as solar and wind, where generation availability varies over time. Within these time-dependent constraints of availability, SBSP is considered fully dispatchable on an hourly basis, allowing its output to respond flexibly to the needs of the system.

At the network level in 2020 and 2050, all generator capacities are set as \textit{extendable}, allowing generation to be governed by cost-efficiency. However, to enhance model realism, we impose specific constraints on different generator types. Due to geographical and environmental limitations, run-of-river and reservoir generators have their maximum capacity \textit{p\_nom\_max} restricted to current installed capacity \textit{p\_nom}, reflecting the limited expansion potential \cite{hydro1, hydro2}. Biomass and nuclear generators are assigned a minimum capacity \textit{p\_nom\_min}  equivalent to their current capacity \textit{p\_nom}, acknowledging the high construction and operational costs of nuclear plants and the sustainable, carbon-neutral attributes of biomass energy \cite{EashGates2020, Deng2024, Ali2024Biomass}. Generators powered by coal, oil, and natural gas, projected to be phased out, have their \textit{p\_nom} set to \textit{0}.

For SBSP generator costs, we draw from modelling outcomes and select cost parameters as outlined in Table \ref{tab:costs}. Using PyPSA’s cost calculation framework, we assign the capital and marginal costs for SBSP generators.

\bmhead{Acknowledgements}

We acknowledge technical discussions with Gabriel Wittenberg. W.H. would like to acknowledge the support of the Royal Academy of Engineering (RAEng) Engineering for Development Research Fellowship [grant number RF\\201819\\18\\89] and UKRI-EPSRC [grant number EP/W027372/1].

\bmhead{Data and codes availability}
All data and codes supporting the findings of this study are available within the paper can be found via \url{https://github.com/CHEperb/SBSP-PyPSA}.

\bibliography{sn-bibliography}


\begin{thebibliography}{76}
\ifx \bisbn   \undefined \def \bisbn  #1{ISBN #1}\fi
\ifx \binits  \undefined \def \binits#1{#1}\fi
\ifx \bauthor  \undefined \def \bauthor#1{#1}\fi
\ifx \batitle  \undefined \def \batitle#1{#1}\fi
\ifx \bjtitle  \undefined \def \bjtitle#1{#1}\fi
\ifx \bvolume  \undefined \def \bvolume#1{\textbf{#1}}\fi
\ifx \byear  \undefined \def \byear#1{#1}\fi
\ifx \bissue  \undefined \def \bissue#1{#1}\fi
\ifx \bfpage  \undefined \def \bfpage#1{#1}\fi
\ifx \blpage  \undefined \def \blpage #1{#1}\fi
\ifx \burl  \undefined \def \burl#1{\textsf{#1}}\fi
\ifx \doiurl  \undefined \def \doiurl#1{\url{https://doi.org/#1}}\fi
\ifx \betal  \undefined \def \betal{\textit{et al.}}\fi
\ifx \binstitute  \undefined \def \binstitute#1{#1}\fi
\ifx \binstitutionaled  \undefined \def \binstitutionaled#1{#1}\fi
\ifx \bctitle  \undefined \def \bctitle#1{#1}\fi
\ifx \beditor  \undefined \def \beditor#1{#1}\fi
\ifx \bpublisher  \undefined \def \bpublisher#1{#1}\fi
\ifx \bbtitle  \undefined \def \bbtitle#1{#1}\fi
\ifx \bedition  \undefined \def \bedition#1{#1}\fi
\ifx \bseriesno  \undefined \def \bseriesno#1{#1}\fi
\ifx \blocation  \undefined \def \blocation#1{#1}\fi
\ifx \bsertitle  \undefined \def \bsertitle#1{#1}\fi
\ifx \bsnm \undefined \def \bsnm#1{#1}\fi
\ifx \bsuffix \undefined \def \bsuffix#1{#1}\fi
\ifx \bparticle \undefined \def \bparticle#1{#1}\fi
\ifx \barticle \undefined \def \barticle#1{#1}\fi
\bibcommenthead
\ifx \bconfdate \undefined \def \bconfdate #1{#1}\fi
\ifx \botherref \undefined \def \botherref #1{#1}\fi
\ifx \url \undefined \def \url#1{\textsf{#1}}\fi
\ifx \bchapter \undefined \def \bchapter#1{#1}\fi
\ifx \bbook \undefined \def \bbook#1{#1}\fi
\ifx \bcomment \undefined \def \bcomment#1{#1}\fi
\ifx \oauthor \undefined \def \oauthor#1{#1}\fi
\ifx \citeauthoryear \undefined \def \citeauthoryear#1{#1}\fi
\ifx \endbibitem  \undefined \def \endbibitem {}\fi
\ifx \bconflocation  \undefined \def \bconflocation#1{#1}\fi
\ifx \arxivurl  \undefined \def \arxivurl#1{\textsf{#1}}\fi
\csname PreBibitemsHook\endcsname

\bibitem[\protect\citeauthoryear{{Energy \& Climate Intelligence Unit and Oxford Net Zero}}{2023}]{netzero_countries}
\begin{botherref}
\oauthor{\bsnm{{Energy \& Climate Intelligence Unit and Oxford Net Zero}}}:
Net Zero Tracker.
Available at \url{https://zerotracker.net/}
(2023)
\end{botherref}
\endbibitem

\bibitem[\protect\citeauthoryear{Schmidt et~al.}{2019}]{schmidt2019projecting}
\begin{barticle}
\bauthor{\bsnm{Schmidt}, \binits{O.}},
\bauthor{\bsnm{Melchior}, \binits{S.}},
\bauthor{\bsnm{Hawkes}, \binits{A.}},
\bauthor{\bsnm{Staffell}, \binits{I.}}:
\batitle{Projecting the future levelized cost of electricity storage technologies}.
\bjtitle{Joule}
\bvolume{3}(\bissue{1}),
\bfpage{81}--\blpage{100}
(\byear{2019})
\end{barticle}
\endbibitem

\bibitem[\protect\citeauthoryear{He et~al.}{2021}]{he2021technologies}
\begin{barticle}
\bauthor{\bsnm{He}, \binits{W.}},
\bauthor{\bsnm{King}, \binits{M.}},
\bauthor{\bsnm{Luo}, \binits{X.}},
\bauthor{\bsnm{Dooner}, \binits{M.}},
\bauthor{\bsnm{Li}, \binits{D.}},
\bauthor{\bsnm{Wang}, \binits{J.}}:
\batitle{Technologies and economics of electric energy storages in power systems: Review and perspective}.
\bjtitle{Advances in Applied Energy}
\bvolume{4},
\bfpage{100060}
(\byear{2021})
\end{barticle}
\endbibitem

\bibitem[\protect\citeauthoryear{Mi{\v{s}}{\'\i}k}{2022}]{mivsik2022eu}
\begin{barticle}
\bauthor{\bsnm{Mi{\v{s}}{\'\i}k}, \binits{M.}}:
\batitle{The eu needs to improve its external energy security}.
\bjtitle{Energy Policy}
\bvolume{165},
\bfpage{112930}
(\byear{2022})
\end{barticle}
\endbibitem

\bibitem[\protect\citeauthoryear{Victoria et~al.}{2020}]{victoria2020early}
\begin{barticle}
\bauthor{\bsnm{Victoria}, \binits{M.}},
\bauthor{\bsnm{Zhu}, \binits{K.}},
\bauthor{\bsnm{Brown}, \binits{T.}},
\bauthor{\bsnm{Andresen}, \binits{G.B.}},
\bauthor{\bsnm{Greiner}, \binits{M.}}:
\batitle{Early decarbonisation of the european energy system pays off}.
\bjtitle{Nature communications}
\bvolume{11}(\bissue{1}),
\bfpage{1}--\blpage{9}
(\byear{2020})
\end{barticle}
\endbibitem

\bibitem[\protect\citeauthoryear{Glaser et~al.}{1974}]{glaser1974feasibility}
\begin{botherref}
\oauthor{\bsnm{Glaser}, \binits{P.E.}},
\oauthor{\bsnm{Maynard}, \binits{O.E.}},
\oauthor{\bsnm{Mackovciak}, \binits{J.}},
\oauthor{\bsnm{Ralph}, \binits{E.}}:
Feasibility study of a satellite solar power station.
Technical report,
NASA
(1974)
\end{botherref}
\endbibitem

\bibitem[\protect\citeauthoryear{Geisz et~al.}{2020}]{geisz2020six}
\begin{barticle}
\bauthor{\bsnm{Geisz}, \binits{J.F.}},
\bauthor{\bsnm{France}, \binits{R.M.}},
\bauthor{\bsnm{Schulte}, \binits{K.L.}},
\bauthor{\bsnm{Steiner}, \binits{M.A.}},
\bauthor{\bsnm{Norman}, \binits{A.G.}},
\bauthor{\bsnm{Guthrey}, \binits{H.L.}},
\bauthor{\bsnm{Young}, \binits{M.R.}},
\bauthor{\bsnm{Song}, \binits{T.}},
\bauthor{\bsnm{Moriarty}, \binits{T.}}:
\batitle{Six-junction iii--v solar cells with 47.1\% conversion efficiency under 143 suns concentration}.
\bjtitle{Nature energy}
\bvolume{5}(\bissue{4}),
\bfpage{326}--\blpage{335}
(\byear{2020})
\end{barticle}
\endbibitem

\bibitem[\protect\citeauthoryear{Lang et~al.}{2020}]{lang2020proton}
\begin{barticle}
\bauthor{\bsnm{Lang}, \binits{F.}},
\bauthor{\bsnm{Jo{\v{s}}t}, \binits{M.}},
\bauthor{\bsnm{Frohna}, \binits{K.}},
\bauthor{\bsnm{K{\"o}hnen}, \binits{E.}},
\bauthor{\bsnm{Al-Ashouri}, \binits{A.}},
\bauthor{\bsnm{Bowman}, \binits{A.R.}},
\bauthor{\bsnm{Bertram}, \binits{T.}},
\bauthor{\bsnm{Morales-Vilches}, \binits{A.B.}},
\bauthor{\bsnm{Koushik}, \binits{D.}},
\bauthor{\bsnm{Tennyson}, \binits{E.M.}}, \betal:
\batitle{Proton radiation hardness of perovskite tandem photovoltaics}.
\bjtitle{Joule}
\bvolume{4}(\bissue{5}),
\bfpage{1054}--\blpage{1069}
(\byear{2020})
\end{barticle}
\endbibitem

\bibitem[\protect\citeauthoryear{}{}]{spiderfab}
\begin{botherref}
\url{https://www.nasa.gov/general/spiderfab-architecture-for-on-orbit-construction-of-kilometer-scale-apertures/}.
Assessed in Nov 2024
\end{botherref}
\endbibitem

\bibitem[\protect\citeauthoryear{Ayling et~al.}{2024}]{ayling2024wireless}
\begin{botherref}
\oauthor{\bsnm{Ayling}, \binits{A.}},
\oauthor{\bsnm{Fikes}, \binits{A.}},
\oauthor{\bsnm{Mizrahi}, \binits{O.S.}},
\oauthor{\bsnm{Wu}, \binits{A.}},
\oauthor{\bsnm{Riazati}, \binits{R.}},
\oauthor{\bsnm{Brunet}, \binits{J.}},
\oauthor{\bsnm{Abiri}, \binits{B.}},
\oauthor{\bsnm{Bohn}, \binits{F.}},
\oauthor{\bsnm{Gal-Katziri}, \binits{M.}},
\oauthor{\bsnm{Hashemi}, \binits{M.R.M.}}, et al.:
Wireless power transfer in space using flexible, lightweight, coherent arrays.
arXiv preprint arXiv:2401.15267
(2024)
\end{botherref}
\endbibitem

\bibitem[\protect\citeauthoryear{Pelham et~al.}{2024}]{pelham2024lyceanem}
\begin{bchapter}
\bauthor{\bsnm{Pelham}, \binits{T.}},
\bauthor{\bsnm{Kudera}, \binits{S.M.}},
\bauthor{\bsnm{Fearon}, \binits{T.C.}}:
\bctitle{Lyceanem: Gigascale electromagnetics for beamforming and system planning}.
In: \bbtitle{International Conference on Energy from Space}
(\byear{2024})
\end{bchapter}
\endbibitem

\bibitem[\protect\citeauthoryear{Jones}{2018}]{jones2018recent}
\begin{bchapter}
\bauthor{\bsnm{Jones}, \binits{H.}}:
\bctitle{The recent large reduction in space launch cost}.
(\byear{2018}).
\bcomment{48th International Conference on Environmental Systems}
\end{bchapter}
\endbibitem

\bibitem[\protect\citeauthoryear{Osoro and Oughton}{2021}]{osoro2021techno}
\begin{barticle}
\bauthor{\bsnm{Osoro}, \binits{O.B.}},
\bauthor{\bsnm{Oughton}, \binits{E.J.}}:
\batitle{A techno-economic framework for satellite networks applied to low earth orbit constellations: Assessing starlink, oneweb and kuiper}.
\bjtitle{IEEE Access}
\bvolume{9},
\bfpage{141611}--\blpage{141625}
(\byear{2021})
\end{barticle}
\endbibitem

\bibitem[\protect\citeauthoryear{Rodgers et~al.}{2024}]{rodgers2024space}
\begin{bchapter}
\bauthor{\bsnm{Rodgers}, \binits{E.}},
\bauthor{\bsnm{Sotudeh}, \binits{J.}},
\bauthor{\bsnm{Mullins}, \binits{C.}},
\bauthor{\bsnm{Hernandez}, \binits{A.}},
\bauthor{\bsnm{Gertsen}, \binits{E.}},
\bauthor{\bsnm{Joseph}, \binits{N.}},
\bauthor{\bsnm{Le}, \binits{H.}},
\bauthor{\bsnm{Smith}, \binits{P.}}:
\bctitle{Space based solar power}.
In: \bbtitle{AIAA AVIATION FORUM AND ASCEND 2024},
p. \bfpage{4944}
(\byear{2024})
\end{bchapter}
\endbibitem

\bibitem[\protect\citeauthoryear{Mizrahi et~al.}{2024}]{mizrahi2024space}
\begin{botherref}
\oauthor{\bsnm{Mizrahi}, \binits{O.}},
\oauthor{\bsnm{Jahelka}, \binits{P.}},
\oauthor{\bsnm{Gdoutos}, \binits{E.}},
\oauthor{\bsnm{Brunet}, \binits{J.}},
\oauthor{\bsnm{Ayling}, \binits{A.}},
\oauthor{\bsnm{Fikes}, \binits{A.}},
\oauthor{\bsnm{Wu}, \binits{A.}},
\oauthor{\bsnm{Madonna}, \binits{R.}},
\oauthor{\bsnm{Atwater}, \binits{H.}},
\oauthor{\bsnm{Pellegrino}, \binits{S.}}, et al.:
Space solar power generation: a viable system proposal and technoeconomic analysis
(2024)
\end{botherref}
\endbibitem

\bibitem[\protect\citeauthoryear{Yang et~al.}{2025}]{yang2025integrated}
\begin{barticle}
\bauthor{\bsnm{Yang}, \binits{C.}},
\bauthor{\bsnm{Wang}, \binits{Q.}},
\bauthor{\bsnm{Lu}, \binits{W.}}, \betal:
\batitle{Integrated uncertain optimal design strategy for truss configuration and attitude--vibration control in rigid--flexible coupling structure with interval uncertainties}.
\bjtitle{Nonlinear Dynamics}
\bvolume{113},
\bfpage{2215}--\blpage{2238}
(\byear{2025})
\doiurl{10.1007/s11071-024-10291-w}
\end{barticle}
\endbibitem

\bibitem[\protect\citeauthoryear{Yang}{2025}]{YANG2025118742}
\begin{barticle}
\bauthor{\bsnm{Yang}, \binits{C.}}:
\batitle{Interval riccati equation-based and non-probabilistic dynamic reliability-constrained multi-objective optimal vibration control with multi-source uncertainties}.
\bjtitle{Journal of Sound and Vibration}
\bvolume{595},
\bfpage{118742}
(\byear{2025})
\doiurl{10.1016/j.jsv.2024.118742}
\end{barticle}
\endbibitem

\bibitem[\protect\citeauthoryear{Yang et~al.}{2025}]{YANG2025115769}
\begin{barticle}
\bauthor{\bsnm{Yang}, \binits{C.}},
\bauthor{\bsnm{Wu}, \binits{J.}},
\bauthor{\bsnm{Fan}, \binits{Z.}},
\bauthor{\bsnm{Lu}, \binits{W.}}:
\batitle{Convex set reliability-based optimal attitude control for space solar power station with bounded and correlated uncertainties}.
\bjtitle{Chaos, Solitons \& Fractals}
\bvolume{190},
\bfpage{115769}
(\byear{2025})
\doiurl{10.1016/j.chaos.2024.115769}
\end{barticle}
\endbibitem

\bibitem[\protect\citeauthoryear{Yang and Liu}{2024}]{YANG20243273}
\begin{barticle}
\bauthor{\bsnm{Yang}, \binits{C.}},
\bauthor{\bsnm{Liu}, \binits{Y.}}:
\batitle{Multi-objective optimization for robust attitude determination of satellite with narrow bound theory}.
\bjtitle{Advances in Space Research}
\bvolume{74}(\bissue{7}),
\bfpage{3273}--\blpage{3283}
(\byear{2024})
\doiurl{10.1016/j.asr.2024.06.002}
\end{barticle}
\endbibitem

\bibitem[\protect\citeauthoryear{}{}]{ESA_sbsp}
\begin{botherref}
Space-Based Solar Power overview.
\url{https://www.esa.int/Enabling_Support/Space_Engineering_Technology/SOLARIS/Space-Based_Solar_Power_overview}.
Accessed in Dec 2024
\end{botherref}
\endbibitem

\bibitem[\protect\citeauthoryear{}{}]{JAXA}
\begin{botherref}
\url{https://www.kenkai.jaxa.jp/eng/research/ssps/ssps-index.html}.
Assessed in Nov 2024
\end{botherref}
\endbibitem

\bibitem[\protect\citeauthoryear{{ENTSO-E and ENTSOG}}{2024}]{TYNDP2024}
\begin{botherref}
\oauthor{\bsnm{{ENTSO-E and ENTSOG}}}:
{TYNDP 2024 Draft Scenarios Report}.
Accessed: 2024-11-10
(2024).
\url{https://2024.entsos-tyndp-scenarios.eu/}
\end{botherref}
\endbibitem

\bibitem[\protect\citeauthoryear{Malaviya et~al.}{2022}]{ref1}
\begin{barticle}
\bauthor{\bsnm{Malaviya}, \binits{P.}},
\bauthor{\bsnm{Sarvaiya}, \binits{V.}},
\bauthor{\bsnm{Shah}, \binits{A.}},
\bauthor{\bsnm{Thakkar}, \binits{D.}},
\bauthor{\bsnm{Shah}, \binits{M.}}:
\batitle{A comprehensive review on space solar power satellite: an idiosyncratic approach}.
\bjtitle{Environmental Science and Pollution Research}
\bvolume{29}(\bissue{28}),
\bfpage{42476}--\blpage{42492}
(\byear{2022})
\end{barticle}
\endbibitem

\bibitem[\protect\citeauthoryear{Jaffe et~al.}{2011}]{ref2}
\begin{bchapter}
\bauthor{\bsnm{Jaffe}, \binits{P.}},
\bauthor{\bsnm{Pasour}, \binits{J.}},
\bauthor{\bsnm{Gonzalez}, \binits{M.}},
\bauthor{\bsnm{Spencer}, \binits{S.}},
\bauthor{\bsnm{Nurnberger}, \binits{M.}},
\bauthor{\bsnm{Dunay}, \binits{J.}},
\bauthor{\bsnm{Scherr}, \binits{M.}},
\bauthor{\bsnm{Jenkins}, \binits{P.}}:
\bctitle{Sandwich module development for space solar power}.
In: \bbtitle{Proceedings of the 28th International Symposium on Space Technology and Science},
pp. \bfpage{5}--\blpage{12}
(\byear{2011})
\end{bchapter}
\endbibitem

\bibitem[\protect\citeauthoryear{Sasaki et~al.}{2007}]{ref3}
\begin{barticle}
\bauthor{\bsnm{Sasaki}, \binits{S.}},
\bauthor{\bsnm{Tanaka}, \binits{K.}},
\bauthor{\bsnm{Higuchi}, \binits{K.}},
\bauthor{\bsnm{Okuizumi}, \binits{N.}},
\bauthor{\bsnm{Kawasaki}, \binits{S.}},
\bauthor{\bsnm{Shinohara}, \binits{N.}},
\bauthor{\bsnm{Senda}, \binits{K.}},
\bauthor{\bsnm{Ishimura}, \binits{K.}}:
\batitle{A new concept of solar power satellite: Tethered-sps}.
\bjtitle{Acta Astronautica}
\bvolume{60}(\bissue{3}),
\bfpage{153}--\blpage{165}
(\byear{2007})
\end{barticle}
\endbibitem

\bibitem[\protect\citeauthoryear{Alam et~al.}{2024}]{ref4}
\begin{botherref}
\oauthor{\bsnm{Alam}, \binits{K.S.}},
\oauthor{\bsnm{Kaif}, \binits{A.D.}},
\oauthor{\bsnm{Das}, \binits{S.K.}},
\oauthor{\bsnm{Abhi}, \binits{S.H.}},
\oauthor{\bsnm{Muyeen}, \binits{S.}},
\oauthor{\bsnm{Ali}, \binits{M.F.}},
\oauthor{\bsnm{Tasneem}, \binits{Z.}},
\oauthor{\bsnm{Islam}, \binits{M.M.}},
\oauthor{\bsnm{Islam}, \binits{M.R.}},
\oauthor{\bsnm{Badal}, \binits{M.F.R.}}, et al.:
Towards net zero: A technological review on the potential of space-based solar power and wireless power transmission.
Heliyon
(2024)
\end{botherref}
\endbibitem

\bibitem[\protect\citeauthoryear{Sasaki et~al.}{2013}]{ref5}
\begin{barticle}
\bauthor{\bsnm{Sasaki}, \binits{S.}},
\bauthor{\bsnm{Tanaka}, \binits{K.}},
\bauthor{\bsnm{Maki}, \binits{K.-i.}}:
\batitle{Microwave power transmission technologies for solar power satellites}.
\bjtitle{Proceedings of the IEEE}
\bvolume{101}(\bissue{6}),
\bfpage{1438}--\blpage{1447}
(\byear{2013})
\end{barticle}
\endbibitem

\bibitem[\protect\citeauthoryear{Osepchuk}{2002}]{ref6}
\begin{barticle}
\bauthor{\bsnm{Osepchuk}, \binits{J.M.}}:
\batitle{How safe are microwaves and solar power from space?}
\bjtitle{IEEE microwave magazine}
\bvolume{3}(\bissue{4}),
\bfpage{58}--\blpage{64}
(\byear{2002})
\end{barticle}
\endbibitem

\bibitem[\protect\citeauthoryear{Mankins}{2021}]{ref13}
\begin{bchapter}
\bauthor{\bsnm{Mankins}, \binits{J.}}:
\bctitle{Sps-alpha mark-iii and an achievable roadmap to space solar power}.
In: \bbtitle{72nd International Astronautical Congress}
(\byear{2021})
\end{bchapter}
\endbibitem

\bibitem[\protect\citeauthoryear{Glaser}{1968}]{ref7}
\begin{barticle}
\bauthor{\bsnm{Glaser}, \binits{P.E.}}:
\batitle{Power from the sun: Its future}.
\bjtitle{Science}
\bvolume{162}(\bissue{3856}),
\bfpage{857}--\blpage{861}
(\byear{1968})
\end{barticle}
\endbibitem

\bibitem[\protect\citeauthoryear{ASSESSMENT}{1980}]{ref8}
\begin{botherref}
\oauthor{\bsnm{ASSESSMENT}, \binits{P.C.}}:
Doe/nasa
(1980)
\end{botherref}
\endbibitem

\bibitem[\protect\citeauthoryear{Mankins}{1997}]{ref9}
\begin{barticle}
\bauthor{\bsnm{Mankins}, \binits{J.C.}}:
\batitle{A fresh look at space solar power: New architectures, concepts and technologies}.
\bjtitle{Acta Astronautica}
\bvolume{41}(\bissue{4-10}),
\bfpage{347}--\blpage{359}
(\byear{1997})
\end{barticle}
\endbibitem

\bibitem[\protect\citeauthoryear{Feingold and Marzwell}{1991}]{Modular}
\begin{bchapter}
\bauthor{\bsnm{Feingold}, \binits{H.}},
\bauthor{\bsnm{Marzwell}, \binits{N.}}:
\bctitle{Modular symmetrical concentrator for space solar power}.
In: \bbtitle{Proceedings of the Space Power Symposium, 42nd Congress of the International Astronautical Federation},
\bconflocation{Montreal, Canada},
pp. \bfpage{123}--\blpage{130}
(\byear{1991})
\end{bchapter}
\endbibitem

\bibitem[\protect\citeauthoryear{Nansen}{2000}]{Arbitrarily}
\begin{barticle}
\bauthor{\bsnm{Nansen}, \binits{R.H.}}:
\batitle{Arbitrarily large phased array for space solar power}.
\bjtitle{Acta Astronautica}
\bvolume{47}(\bissue{2-9}),
\bfpage{575}--\blpage{583}
(\byear{2000})
\doiurl{10.1016/S0094-5765(00)00085-7}
\end{barticle}
\endbibitem

\bibitem[\protect\citeauthoryear{Rohatgi and von Bieren}{1997}]{reflectors}
\begin{barticle}
\bauthor{\bsnm{Rohatgi}, \binits{A.}},
\bauthor{\bsnm{Bieren}, \binits{P.}}:
\batitle{Space-based reflectors for ground-based solar power generation}.
\bjtitle{Solar Energy}
\bvolume{59}(\bissue{4-6}),
\bfpage{115}--\blpage{120}
(\byear{1997})
\doiurl{10.1016/S0038-092X(97)00089-4}
\end{barticle}
\endbibitem

\bibitem[\protect\citeauthoryear{Mankins and Strait}{2000}]{thin-film}
\begin{barticle}
\bauthor{\bsnm{Mankins}, \binits{J.C.}},
\bauthor{\bsnm{Strait}, \binits{M.M.}}:
\batitle{Inflatable thin-film structures for space solar power}.
\bjtitle{Space Technology}
\bvolume{20}(\bissue{2}),
\bfpage{57}--\blpage{63}
(\byear{2000})
\end{barticle}
\endbibitem

\bibitem[\protect\citeauthoryear{Mankins}{2002}]{Hypermodular}
\begin{bchapter}
\bauthor{\bsnm{Mankins}, \binits{J.C.}}:
\bctitle{Hypermodular space solar power design: A fresh approach to space solar power}.
In: \bbtitle{Proceedings of the 53rd International Astronautical Congress},
\bconflocation{Houston, Texas, USA}
(\byear{2002})
\end{bchapter}
\endbibitem

\bibitem[\protect\citeauthoryear{Hoyt and Forward}{1995}]{Tethered}
\begin{bchapter}
\bauthor{\bsnm{Hoyt}, \binits{R.P.}},
\bauthor{\bsnm{Forward}, \binits{R.L.}}:
\bctitle{Tethered satellite system for space-based solar power}.
In: \bbtitle{Proceedings of the 45th International Astronautical Congress},
\bconflocation{Oslo, Norway}
(\byear{1995})
\end{bchapter}
\endbibitem

\bibitem[\protect\citeauthoryear{Rodgers et~al.}{2024}]{ref11}
\begin{bchapter}
\bauthor{\bsnm{Rodgers}, \binits{E.}},
\bauthor{\bsnm{Sotudeh}, \binits{J.}},
\bauthor{\bsnm{Mullins}, \binits{C.}},
\bauthor{\bsnm{Hernandez}, \binits{A.}},
\bauthor{\bsnm{Gertsen}, \binits{E.}},
\bauthor{\bsnm{Joseph}, \binits{N.}},
\bauthor{\bsnm{Le}, \binits{H.}},
\bauthor{\bsnm{Smith}, \binits{P.}}:
\bctitle{Space based solar power}.
In: \bbtitle{AIAA AVIATION FORUM AND ASCEND 2024},
p. \bfpage{4944}
(\byear{2024})
\end{bchapter}
\endbibitem

\bibitem[\protect\citeauthoryear{Mankins et~al.}{2012}]{ref10}
\begin{bchapter}
\bauthor{\bsnm{Mankins}, \binits{J.}},
\bauthor{\bsnm{Kaya}, \binits{N.}},
\bauthor{\bsnm{Vasile}, \binits{M.}}:
\bctitle{Sps-alpha: the first practical solar power satellite via arbitrarily large phased array (a 2011-2012 niac project)}.
In: \bbtitle{10th International Energy Conversion Engineering Conference},
p. \bfpage{3978}
(\byear{2012})
\end{bchapter}
\endbibitem

\bibitem[\protect\citeauthoryear{Bucknell}{2024}]{ref14}
\begin{botherref}
\oauthor{\bsnm{Bucknell}, \binits{J.}}:
Survey of Space Based Solar Power (SBSP).
\url{https://www.researchgate.net/publication/377865839_Survey_of_Space_Based_Solar_Power_SBSP}
(2024)
\end{botherref}
\endbibitem

\bibitem[\protect\citeauthoryear{Fikes et~al.}{2022}]{ref12}
\begin{bchapter}
\bauthor{\bsnm{Fikes}, \binits{A.}},
\bauthor{\bsnm{Gal-Karziri}, \binits{M.}},
\bauthor{\bsnm{Gdoutos}, \binits{E.}},
\bauthor{\bsnm{Kelzenberg}, \binits{M.}},
\bauthor{\bsnm{Warmann}, \binits{E.}},
\bauthor{\bsnm{Madonna}, \binits{R.}},
\bauthor{\bsnm{Atwater}, \binits{H.}},
\bauthor{\bsnm{Hajimiri}, \binits{A.}},
\bauthor{\bsnm{Pellegrino}, \binits{S.}}:
\bctitle{The caltech space solar power project: Design, progress, and future direction}.
In: \bbtitle{Proc. IEEE WiSEE Space Sol. Power Workshop},
pp. \bfpage{1}--\blpage{5}
(\byear{2022})
\end{bchapter}
\endbibitem

\bibitem[\protect\citeauthoryear{Gurnett}{2009}]{gurnett2009search}
\begin{barticle}
\bauthor{\bsnm{Gurnett}, \binits{D.A.}}:
\batitle{The search for life in the solar system}.
\bjtitle{Transactions of the American Clinical and Climatological Association}
\bvolume{120},
\bfpage{299}--\blpage{325}
(\byear{2009})
\end{barticle}
\endbibitem

\bibitem[\protect\citeauthoryear{{Lazard}}{2023}]{Lazard2023LCOE}
\begin{botherref}
\oauthor{\bsnm{{Lazard}}}:
{Lazard's Levelized Cost of Energy+ April 2023}.
Technical report,
Lazard
(2023).
{Accessed: October 26, 2024}.
\url{https://www.lazard.com/media/2ozoovyg/lazards-lcoeplus-april-2023.pdf}
\end{botherref}
\endbibitem

\bibitem[\protect\citeauthoryear{{NREL}}{2022}]{nrel2022definitions}
\begin{botherref}
\oauthor{\bsnm{{NREL}}}:
Definitions.
Accessed: 2025-02-26
(2022).
\url{https://atb.nrel.gov/electricity/2022/definitions#capex}
\end{botherref}
\endbibitem

\bibitem[\protect\citeauthoryear{H{\"o}rsch et~al.}{2018}]{horsch2018pypsa}
\begin{barticle}
\bauthor{\bsnm{H{\"o}rsch}, \binits{J.}},
\bauthor{\bsnm{Hofmann}, \binits{F.}},
\bauthor{\bsnm{Schlachtberger}, \binits{D.}},
\bauthor{\bsnm{Brown}, \binits{T.}}:
\batitle{Pypsa-eur: An open optimisation model of the european transmission system}.
\bjtitle{Energy strategy reviews}
\bvolume{22},
\bfpage{207}--\blpage{215}
(\byear{2018})
\end{barticle}
\endbibitem

\bibitem[\protect\citeauthoryear{Yang et~al.}{2025a}]{YANG2025118929}
\begin{barticle}
\bauthor{\bsnm{Yang}, \binits{C.}},
\bauthor{\bsnm{Xu}, \binits{X.}},
\bauthor{\bsnm{Wang}, \binits{X.}},
\bauthor{\bsnm{Fan}, \binits{Z.}}:
\batitle{Interval-oriented eigensystem realization algorithm and its modification for structural modal parameter identification with bounded uncertainties}.
\bjtitle{Journal of Sound and Vibration}
\bvolume{601},
\bfpage{118929}
(\byear{2025})
\doiurl{10.1016/j.jsv.2025.118929}
\end{barticle}
\endbibitem

\bibitem[\protect\citeauthoryear{Yang et~al.}{2025b}]{10843835}
\begin{botherref}
\oauthor{\bsnm{Yang}, \binits{C.}},
\oauthor{\bsnm{Liu}, \binits{Y.}},
\oauthor{\bsnm{Gao}, \binits{H.}}:
Reliability-constrained uncertain spacecraft sliding mode attitude tracking control with interval parameters.
IEEE Transactions on Aerospace and Electronic Systems,
1--14
(2025)
\doiurl{10.1109/TAES.2025.3529798}
\end{botherref}
\endbibitem

\bibitem[\protect\citeauthoryear{Yang and Yu}{2025}]{YANG2025117814}
\begin{barticle}
\bauthor{\bsnm{Yang}, \binits{C.}},
\bauthor{\bsnm{Yu}, \binits{Q.}}:
\batitle{Multi-objective optimization-inspired set theory-based regularization approach for force reconstruction with bounded uncertainties}.
\bjtitle{Computer Methods in Applied Mechanics and Engineering}
\bvolume{438},
\bfpage{117814}
(\byear{2025})
\doiurl{10.1016/j.cma.2025.117814}
\end{barticle}
\endbibitem

\bibitem[\protect\citeauthoryear{Choi et~al.}{2024}]{CHOI202427}
\begin{barticle}
\bauthor{\bsnm{Choi}, \binits{J.-M.}},
\bauthor{\bsnm{Choi}, \binits{S.-J.}},
\bauthor{\bsnm{Yi}, \binits{S.-H.}}:
\batitle{Case studies on space solar power in korea}.
\bjtitle{Space Solar Power and Wireless Transmission}
\bvolume{1}(\bissue{1}),
\bfpage{27}--\blpage{36}
(\byear{2024})
\doiurl{10.1016/j.sspwt.2024.03.001}
\end{barticle}
\endbibitem

\bibitem[\protect\citeauthoryear{{International Atomic Energy Agency}}{2021}]{IAEA2021}
\begin{bbook}
\bauthor{\bsnm{{International Atomic Energy Agency}}}:
\bbtitle{Technology Roadmap for Small Modular Reactor Deployment}.
\bsertitle{IAEA Nuclear Energy Series},
vol. \bseriesno{NR-T-1.18}.
\bpublisher{International Atomic Energy Agency},
\blocation{Vienna}
(\byear{2021})
\end{bbook}
\endbibitem

\bibitem[\protect\citeauthoryear{{World Nuclear News}}{2021}]{WNN2021HTRPM}
\begin{botherref}
\oauthor{\bsnm{{World Nuclear News}}}:
Demonstration HTR-PM Connected to Grid.
Accessed: 2025-05-09.
\url{https://www.world-nuclear-news.org/Articles/Demonstration-HTR-PM-connected-to-grid}
\end{botherref}
\endbibitem

\bibitem[\protect\citeauthoryear{{International Energy Agency}}{2024}]{IEA2024Hydrogen}
\begin{botherref}
\oauthor{\bsnm{{International Energy Agency}}}:
Hydrogen – Tracking Clean Energy Progress.
International Energy Agency. Accessed: 2025-05-09.
\url{https://www.iea.org/energy-system/low-emission-fuels/hydrogen}
\end{botherref}
\endbibitem

\bibitem[\protect\citeauthoryear{He et~al.}{2021}]{HE2021100060}
\begin{barticle}
\bauthor{\bsnm{He}, \binits{W.}},
\bauthor{\bsnm{King}, \binits{M.}},
\bauthor{\bsnm{Luo}, \binits{X.}},
\bauthor{\bsnm{Dooner}, \binits{M.}},
\bauthor{\bsnm{Li}, \binits{D.}},
\bauthor{\bsnm{Wang}, \binits{J.}}:
\batitle{Technologies and economics of electric energy storages in power systems: Review and perspective}.
\bjtitle{Advances in Applied Energy}
\bvolume{4},
\bfpage{100060}
(\byear{2021})
\doiurl{10.1016/j.adapen.2021.100060}
\end{barticle}
\endbibitem

\bibitem[\protect\citeauthoryear{Sepulveda et~al.}{2018}]{SEPULVEDA20182403}
\begin{barticle}
\bauthor{\bsnm{Sepulveda}, \binits{N.A.}},
\bauthor{\bsnm{Jenkins}, \binits{J.D.}},
\bauthor{\bsnm{{de Sisternes}}, \binits{F.J.}},
\bauthor{\bsnm{Lester}, \binits{R.K.}}:
\batitle{The role of firm low-carbon electricity resources in deep decarbonization of power generation}.
\bjtitle{Joule}
\bvolume{2}(\bissue{11}),
\bfpage{2403}--\blpage{2420}
(\byear{2018})
\doiurl{10.1016/j.joule.2018.08.006}
\end{barticle}
\endbibitem

\bibitem[\protect\citeauthoryear{Hoersch et~al.}{2018}]{PyPSAEur}
\begin{barticle}
\bauthor{\bsnm{Hoersch}, \binits{J.}},
\bauthor{\bsnm{Hofmann}, \binits{F.}},
\bauthor{\bsnm{Schlachtberger}, \binits{D.}},
\bauthor{\bsnm{Brown}, \binits{T.}}:
\batitle{Pypsa-eur: An open optimisation model of the european transmission system}.
\bjtitle{Energy Strategy Reviews}
\bvolume{22},
\bfpage{207}--\blpage{215}
(\byear{2018})
\doiurl{10.1016/j.esr.2018.08.012}
{\href{https://arxiv.org/abs/1806.01613}{{1806.01613}}}
\end{barticle}
\endbibitem

\bibitem[\protect\citeauthoryear{Sasse and Trutnevyte}{2020}]{sasse2020regional}
\begin{barticle}
\bauthor{\bsnm{Sasse}, \binits{J.}},
\bauthor{\bsnm{Trutnevyte}, \binits{E.}}:
\batitle{Regional impacts of electricity system transition in central europe until 2035}.
\bjtitle{Nature Communications}
\bvolume{11},
\bfpage{4972}
(\byear{2020})
\doiurl{10.1038/s41467-020-18812-y}
\end{barticle}
\endbibitem

\bibitem[\protect\citeauthoryear{Brown et~al.}{2018}]{PyPSA}
\begin{botherref}
\oauthor{\bsnm{Brown}, \binits{T.}},
\oauthor{\bsnm{H\"orsch}, \binits{J.}},
\oauthor{\bsnm{Schlachtberger}, \binits{D.}}:
{PyPSA: Python for Power System Analysis}.
Journal of Open Research Software
\textbf{6}(4)
(2018)
\doiurl{10.5334/jors.188}
{\href{https://arxiv.org/abs/1707.09913}{{1707.09913}}}
\end{botherref}
\endbibitem

\bibitem[\protect\citeauthoryear{{ENTSO-E}}{}]{entsoe_load_data}
\begin{botherref}
\oauthor{\bsnm{{ENTSO-E}}}:
Total Load Data.
\url{https://transparency.entsoe.eu/load-domain/r2/totalLoadR2/show}.
Accessed: 2024-09-23
\end{botherref}
\endbibitem

\bibitem[\protect\citeauthoryear{Gotzens et~al.}{2019}]{gotzens_performing_2019}
\begin{barticle}
\bauthor{\bsnm{Gotzens}, \binits{F.}},
\bauthor{\bsnm{Heinrichs}, \binits{H.}},
\bauthor{\bsnm{Hörsch}, \binits{J.}},
\bauthor{\bsnm{Hofmann}, \binits{F.}}:
\batitle{Performing energy modelling exercises in a transparent way - {The} issue of data quality in power plant databases}.
\bjtitle{Energy Strategy Reviews}
\bvolume{23},
\bfpage{1}--\blpage{12}
(\byear{2019})
\doiurl{10.1016/j.esr.2018.11.004} .
Accessed 2018-12-03
\end{barticle}
\endbibitem

\bibitem[\protect\citeauthoryear{Hersbach et~al.}{2020}]{https://doi.org/10.1002/qj.3803}
\begin{barticle}
\bauthor{\bsnm{Hersbach}, \binits{H.}},
\bauthor{\bsnm{Bell}, \binits{B.}},
\bauthor{\bsnm{Berrisford}, \binits{P.}},
\bauthor{\bsnm{Hirahara}, \binits{S.}},
\bauthor{\bsnm{Horányi}, \binits{A.}},
\bauthor{\bsnm{Muñoz-Sabater}, \binits{J.}},
\bauthor{\bsnm{Nicolas}, \binits{J.}},
\bauthor{\bsnm{Peubey}, \binits{C.}},
\bauthor{\bsnm{Radu}, \binits{R.}},
\bauthor{\bsnm{Schepers}, \binits{D.}},
\bauthor{\bsnm{Simmons}, \binits{A.}},
\bauthor{\bsnm{Soci}, \binits{C.}},
\bauthor{\bsnm{Abdalla}, \binits{S.}},
\bauthor{\bsnm{Abellan}, \binits{X.}},
\bauthor{\bsnm{Balsamo}, \binits{G.}},
\bauthor{\bsnm{Bechtold}, \binits{P.}},
\bauthor{\bsnm{Biavati}, \binits{G.}},
\bauthor{\bsnm{Bidlot}, \binits{J.}},
\bauthor{\bsnm{Bonavita}, \binits{M.}},
\bauthor{\bsnm{De~Chiara}, \binits{G.}},
\bauthor{\bsnm{Dahlgren}, \binits{P.}},
\bauthor{\bsnm{Dee}, \binits{D.}},
\bauthor{\bsnm{Diamantakis}, \binits{M.}},
\bauthor{\bsnm{Dragani}, \binits{R.}},
\bauthor{\bsnm{Flemming}, \binits{J.}},
\bauthor{\bsnm{Forbes}, \binits{R.}},
\bauthor{\bsnm{Fuentes}, \binits{M.}},
\bauthor{\bsnm{Geer}, \binits{A.}},
\bauthor{\bsnm{Haimberger}, \binits{L.}},
\bauthor{\bsnm{Healy}, \binits{S.}},
\bauthor{\bsnm{Hogan}, \binits{R.J.}},
\bauthor{\bsnm{Hólm}, \binits{E.}},
\bauthor{\bsnm{Janisková}, \binits{M.}},
\bauthor{\bsnm{Keeley}, \binits{S.}},
\bauthor{\bsnm{Laloyaux}, \binits{P.}},
\bauthor{\bsnm{Lopez}, \binits{P.}},
\bauthor{\bsnm{Lupu}, \binits{C.}},
\bauthor{\bsnm{Radnoti}, \binits{G.}},
\bauthor{\bsnm{Rosnay}, \binits{P.}},
\bauthor{\bsnm{Rozum}, \binits{I.}},
\bauthor{\bsnm{Vamborg}, \binits{F.}},
\bauthor{\bsnm{Villaume}, \binits{S.}},
\bauthor{\bsnm{Thépaut}, \binits{J.-N.}}:
\batitle{The era5 global reanalysis}.
\bjtitle{Quarterly Journal of the Royal Meteorological Society}
\bvolume{146}(\bissue{730}),
\bfpage{1999}--\blpage{2049}
(\byear{2020})
\doiurl{10.1002/qj.3803}
{\href{https://arxiv.org/abs/https://rmets.onlinelibrary.wiley.com/doi/pdf/10.1002/qj.3803}{{https://rmets.onlinelibrary.wiley.com/doi/pdf/10.1002/qj.3803}}}
\end{barticle}
\endbibitem

\bibitem[\protect\citeauthoryear{{Danish Energy Agency}}{2020}]{DEA2020}
\begin{botherref}
\oauthor{\bsnm{{Danish Energy Agency}}}:
{Technology Data}.
{Accessed: [date you accessed the website]}
(2020).
\url{https://ens.dk/en/our-services/projections-and-models/technology-data}
\end{botherref}
\endbibitem

\bibitem[\protect\citeauthoryear{{International Energy Agency}}{2023}]{IEAGlobal2023}
\begin{botherref}
\oauthor{\bsnm{{International Energy Agency}}}:
{Global Energy and Climate Model 2023 Key Input Data}.
{Licence: Terms of Use for Non-CC Material}.
{IEA, Paris}
(2023).
\url{https://www.iea.org/data-and-statistics/data-product/global-energy-and-climate-model-2023-key-input-data}
\end{botherref}
\endbibitem

\bibitem[\protect\citeauthoryear{lisazeyen et~al.}{2024}]{PyPSA2024}
\begin{botherref}
\oauthor{\bsnm{lisazeyen}},
\oauthor{\bsnm{euronion}},
\oauthor{\bsnm{Neumann}, \binits{F.}},
\oauthor{\bsnm{Millinger}, \binits{M.}},
\oauthor{\bsnm{Parzen}, \binits{M.}},
\oauthor{\bsnm{aalamia}},
\oauthor{\bsnm{Franken}, \binits{L.}},
\oauthor{\bsnm{Brown}, \binits{T.}},
\oauthor{\bsnm{Geis}, \binits{J.}},
\oauthor{\bsnm{Glaum}, \binits{P.}},
\oauthor{\bsnm{martavp}},
\oauthor{\bsnm{cpschau}},
\oauthor{\bsnm{Greevenbroek}, \binits{K.}},
\oauthor{\bsnm{Trippe}, \binits{L.}},
\oauthor{\bsnm{fhg-isi}},
\oauthor{\bsnm{lukasnacken}},
\oauthor{\bsnm{s8au}},
\oauthor{\bsnm{Seibold}, \binits{T.}}:
{PyPSA/technology-data: v0.9.2}
(2024).
\doiurl{10.5281/zenodo.13617294} .
\url{https://doi.org/10.5281/zenodo.13617294}
\end{botherref}
\endbibitem

\bibitem[\protect\citeauthoryear{{European Commission}}{2021a}]{EU2021carbon}
\begin{botherref}
\oauthor{\bsnm{{European Commission}}}:
{Report from the Commission to the European Parliament and the Council on the Functioning of the European Carbon Market in 2020}.
{COM(2021) 962 final}
(2021).
\url{https://climate.ec.europa.eu/document/download/abe3f38e-b4d0-4c4b-9459-9f05d618b493_en}
\end{botherref}
\endbibitem

\bibitem[\protect\citeauthoryear{{European Commission}}{2021b}]{EUClimateLaw2021}
\begin{botherref}
\oauthor{\bsnm{{European Commission}}}:
{European Climate Law}.
{Accessed: [your access date]}
(2021).
\url{https://climate.ec.europa.eu/eu-action/european-climate-law_en}
\end{botherref}
\endbibitem

\bibitem[\protect\citeauthoryear{Unnewehr et~al.}{2022}]{validation1}
\begin{bchapter}
\bauthor{\bsnm{Unnewehr}, \binits{J.F.}},
\bauthor{\bsnm{Schäfer}, \binits{M.}},
\bauthor{\bsnm{Weidlich}, \binits{A.}}:
\bctitle{The value of network resolution – a validation study of the european energy system model pypsa-eur}.
In: \bbtitle{2022 Open Source Modelling and Simulation of Energy Systems (OSMSES)},
pp. \bfpage{1}--\blpage{7}
(\byear{2022}).
\doiurl{10.1109/OSMSES54027.2022.9769123}
\end{bchapter}
\endbibitem

\bibitem[\protect\citeauthoryear{Hofmann et~al.}{2021}]{Hofmann2021}
\begin{barticle}
\bauthor{\bsnm{Hofmann}, \binits{F.}},
\bauthor{\bsnm{Hampp}, \binits{J.}},
\bauthor{\bsnm{Neumann}, \binits{F.}},
\bauthor{\bsnm{Brown}, \binits{T.}},
\bauthor{\bsnm{Hörsch}, \binits{J.}}:
\batitle{atlite: A lightweight python package for calculating renewable power potentials and time series}.
\bjtitle{Journal of Open Source Software}
\bvolume{6}(\bissue{62}),
\bfpage{3294}
(\byear{2021})
\doiurl{10.21105/joss.03294}
\end{barticle}
\endbibitem

\bibitem[\protect\citeauthoryear{{ENTSO-E}}{}]{entsoe_actual_generation}
\begin{botherref}
\oauthor{\bsnm{{ENTSO-E}}}:
Actual Generation per Production Type.
\url{https://transparency.entsoe.eu/generation/r2/actualGenerationPerProductionType/show}.
Accessed: 2024-10-20
\end{botherref}
\endbibitem

\bibitem[\protect\citeauthoryear{{ENTSO-E}}{}]{entsoe_generation_capacity}
\begin{botherref}
\oauthor{\bsnm{{ENTSO-E}}}:
Installed Generation Capacity Aggregation.
\url{https://transparency.entsoe.eu/generation/r2/installedGenerationCapacityAggregation/show}.
Accessed: 2024-10-20
\end{botherref}
\endbibitem

\bibitem[\protect\citeauthoryear{Wilczak et~al.}{2024}]{era5}
\begin{botherref}
\oauthor{\bsnm{Wilczak}, \binits{J.M.}},
\oauthor{\bsnm{Akish}, \binits{E.}},
\oauthor{\bsnm{Capotondi}, \binits{A.}},
\oauthor{\bsnm{Compo}, \binits{G.P.}}:
Evaluation and bias correction of the era5 reanalysis over the united states for wind and solar energy applications.
Energies
\textbf{17}(7)
(2024)
\doiurl{10.3390/en17071667}
\end{botherref}
\endbibitem

\bibitem[\protect\citeauthoryear{Kuriqi et~al.}{2021}]{hydro1}
\begin{barticle}
\bauthor{\bsnm{Kuriqi}, \binits{A.}},
\bauthor{\bsnm{Pinheiro}, \binits{A.}},
\bauthor{\bsnm{Sordo-Ward}, \binits{A.}},
\bauthor{\bsnm{Bejarano}, \binits{M.}},
\bauthor{\bsnm{Garrote}, \binits{L.}}:
\batitle{Ecological impacts of run-of-river hydropower plants-current status and future prospects on the brink of energy transition}.
\bjtitle{Renewable and Sustainable Energy Reviews}
\bvolume{142},
\bfpage{110833}
(\byear{2021})
\doiurl{10.1016/j.rser.2021.110833}
\end{barticle}
\endbibitem

\bibitem[\protect\citeauthoryear{Zarfl et~al.}{2015}]{hydro2}
\begin{barticle}
\bauthor{\bsnm{Zarfl}, \binits{C.}},
\bauthor{\bsnm{Lumsdon}, \binits{A.E.}},
\bauthor{\bsnm{Berlekamp}, \binits{J.}},
\bauthor{\bsnm{Tydecks}, \binits{L.}},
\bauthor{\bsnm{Tockner}, \binits{K.}}:
\batitle{A global boom in hydropower dam construction}.
\bjtitle{Aquatic Sciences}
\bvolume{77},
\bfpage{161}--\blpage{170}
(\byear{2015})
\doiurl{10.1007/s00027-014-0377-0}
\end{barticle}
\endbibitem

\bibitem[\protect\citeauthoryear{Eash-Gates et~al.}{2020}]{EashGates2020}
\begin{barticle}
\bauthor{\bsnm{Eash-Gates}, \binits{P.}},
\bauthor{\bsnm{Buongiorno}, \binits{J.}},
\bauthor{\bsnm{Corradini}, \binits{M.}},
\bauthor{\bsnm{Parsons}, \binits{J.E.}}:
\batitle{Sources of cost overrun in nuclear power plant construction call for a new approach to engineering design}.
\bjtitle{Joule}
\bvolume{4}(\bissue{11}),
\bfpage{2348}--\blpage{2373}
(\byear{2020})
\doiurl{10.1016/j.joule.2020.09.017}
\end{barticle}
\endbibitem

\bibitem[\protect\citeauthoryear{Deng et~al.}{2024}]{Deng2024}
\begin{barticle}
\bauthor{\bsnm{Deng}, \binits{X.}},
\bauthor{\bsnm{Teng}, \binits{F.}},
\bauthor{\bsnm{Chen}, \binits{M.}}, \betal:
\batitle{Exploring negative emission potential of biochar to achieve carbon neutrality goal in china}.
\bjtitle{Nature Communications}
\bvolume{15},
\bfpage{1085}
(\byear{2024})
\doiurl{10.1038/s41467-024-45314-y}
\end{barticle}
\endbibitem

\bibitem[\protect\citeauthoryear{Ali et~al.}{2024}]{Ali2024Biomass}
\begin{barticle}
\bauthor{\bsnm{Ali}, \binits{F.}},
\bauthor{\bsnm{Dawood}, \binits{A.}},
\bauthor{\bsnm{Hussain}, \binits{A.}}, \betal:
\batitle{Fueling the future: biomass applications for green and sustainable energy}.
\bjtitle{Discover Sustainability}
\bvolume{5},
\bfpage{156}
(\byear{2024})
\doiurl{10.1007/s43621-024-00309-z}
\end{barticle}
\endbibitem

\end{thebibliography}

\end{document}